\RequirePackage{rotating}

\documentclass[fleqn,usenatbib]{mnras}
\usepackage{newtxtext,newtxmath}
\usepackage{txfonts}
\usepackage[T1]{fontenc}

\usepackage{upgreek}

\DeclareRobustCommand{\VAN}[3]{#2}
\let\VANthebibliography\thebibliography
\def\thebibliography{\DeclareRobustCommand{\VAN}[3]{##3}\VANthebibliography}

\usepackage{graphicx}	
\usepackage{amsmath}	
\usepackage{rotating}

\title[Optical spectroscopy of blazars for CTA -- II]{Optical Spectroscopy of Blazars for the Cherenkov Telescope Array -- II}

\author[E. Kasai et al.]{E. Kasai$^{1}$\thanks{ekasai@unam.na},
P. Goldoni$^{2}$,
S. Pita$^{3}$,
D. A. Williams$^4$,
W. Max-Moerbeck$^{5}$,
O. Hervet$^{4}$,
G. Cotter$^{6}$,\newauthor
M. Backes$^{1,7},$
C. Boisson$^{8},$
J. Becerra González$^{9,10}$,
U. Barres de Almeida$^{11}$,
F. D'Ammando$^{12}$,\newauthor
V. Fallah Ramazani$^{13,14}$, 
E. Lindfors$^{13}$
\\
$^{1}$Department of Physics, Chemistry \& Material Science, University of Namibia, Private Bag 13301, Windhoek, Namibia\\
$^{2}$Universit\'e Paris Cit\'e, CNRS, CEA, Astroparticule et Cosmologie, F-75013 Paris, France\\
$^{3}$Universit\'e Paris Cit\'e, CNRS, Astroparticule et Cosmologie, F-75013 Paris, France\\
$^{4}$Santa Cruz Institute for Particle Physics and Department of Physics, University of California, Santa Cruz, Santa Cruz, CA \\
$^{5}$Departamento de Astronomía, Universidad de Chile, Camino El Observatorio 1515, Las Condes, Santiago, Chile\\
$^{6}$University of Oxford, Oxford Astrophysics, Denys Wilkinson Building, Keble Road, Oxford, OX1 3RH, United Kingdom\\
$^{7}$Centre for Space Research, North-West University, Potchefstroom 2520, South Africa\\
$^{8}$LUTH, Observatoire de Paris,  PSL Research University, CNRS,Universit\'{e} Paris Diderot, Meudon, France\\
$^{9}$Universidad de La Laguna (ULL), Departamento de Astrof\'isica, E-38206 La Laguna, Tenerife, Spain\\
$^{10}$Instituto de Astrof\'isica de Canarias (IAC), E-38200 La Laguna, Tenerife, Spain\\
$^{11}$Centro Brasileiro de Pesquisas Físicas (CBPF), Rua Dr. Xavier Sigaud 150, 22290-180 Rio de Janeiro, Brazil\\
$^{12}$INAF - Istituto di Radioastronomia, Via Gobetti 101, I-40129 Bologna, Italy\\
$^{13}$Finnish Centre for Astronomy with ESO (FINCA), Quantum, Vesilinnantie 5, FI-20014, University of Turku, Finland\\
$^{14}$Ruhr-Universit\"at Bochum, Fakult\"at f\"ur Physik und Astronomie, Astronomisches Institut (AIRUB), 44801 Bochum, Germany\\
}

\date{Accepted 2022 October 24. Received 2022 October 24; in original form 2022 August 24}

\pubyear{2022}

\begin{document}
\label{firstpage}
\pagerange{\pageref{firstpage}--\pageref{lastpage}}
\maketitle

\begin{abstract}
 {Active galactic nuclei (AGNs) make up about 35 per cent of the more than 250 sources detected in very-high-energy (VHE) gamma rays to date with Imaging Atmospheric Cherenkov Telescopes. Apart from four nearby radio galaxies and two AGNs of unknown type, all known VHE AGNs are blazars. Knowledge of the cosmological redshift of gamma-ray blazars is key to enabling the study of their intrinsic emission properties, as the interaction between gamma rays
and the extragalactic background light (EBL) results in a spectral softening. Therefore, the redshift determination exercise is crucial to indirectly placing tight constraints on the EBL density and to studying blazar population evolution across cosmic time. Due to the powerful relativistic jets in blazars, most of their host galaxies’ spectral features are outshined, and dedicated high
signal-to-noise spectroscopic observations are required. Deep medium- to high-resolution spectroscopy of 33 gamma-ray blazar optical counterparts was performed with the European Southern Observatory New Technology Telescope, Keck II telescope, Shane 3-meter telescope and the Southern African Large Telescope. From the sample, spectra from 25 objects display spectral features or are featureless and have high signal-to-noise. The other eight objects have low quality featureless spectra. We systematically searched for absorption and emission features and estimated, when possible, the fractional host galaxy flux in the measured total flux. Our measurements yielded 14 firm spectroscopic redshifts, ranging from 0.0838 to 0.8125, one tentative redshift, and two lower limits: one at z~>~0.382 and the other at z > 0.629.}
\end{abstract}




\section{Introduction}
Studies of AGNs provide insights into the cosmological evolution of star and galaxy formation through various methods, including constraints on photon
fields \citep{Gould1967} and magnetic fields \citep{Aha94,Alves19} along the line of sight from Earth. Such studies also advance our understanding of the physics of black holes and their accretion mechanisms \citep{Scta19}, both of which remain active areas of research in high-energy~(HE,~$E$~>~100 MeV) and VHE ($E$~>~100~GeV) astrophysics. Of the more than 250 sources (including the two starburst galaxies M 82 and NGC 253) detected in the VHE band to date with Imaging  telescopes\footnote{\url{http://tevcat.uchicago.edu}}, about 35~per cent are AGNs and all are classified as blazars, with the exception of four nearby radio galaxies. Blazars are a subclass of the radio-loud AGNs \citep{kellerman89}--which possess powerful relativistic jets, i.e. beamed outflows of relativistic particles originating near the active nucleus \citep{Beall15, Blandford19}--and their jets are aligned nearly along the line of sight to the observer. Of all types of AGNs, blazars show the most extreme observational characteristics. Their radiation is dominated by non-thermal emission, spanning the entire electromagnetic (EM) spectrum with strong polarization in the radio \citep{Ledden78, Lis11} and optical \citep[][]{Angel80, Angelakis16} regimes, and observed to vary across all frequencies on timescales of years, months, days and minutes \citep{Bignami81, Schreier82, Punch92, Montigny95, Aharo07}. In the literature, such observational variabilities are understood to be caused by the jet emission undergoing strong Doppler amplification \citep[see e.g.][]{jorstad17}.

The spectral energy distribution (SED) of blazars has a characteristic broad, double-peaked shape that consists of low- and high-energy components. The low-energy component ranges from the radio to the visible or UV/X-ray bands and is believed to be caused by synchrotron emission from ultrarelativistic electrons accelerated in the jet. The physical processes responsible for the high-energy peak are not unambiguously established. Different candidate processes have been proposed within the framework of leptonic and hadronic scenarios. The leptonic models postulate that the emission is due to soft photons undergoing Compton upscattering by (1) the same electrons which emit the synchrotron radiation (synchrotron self-Compton models) and (2) electrons that are not necessarily the same as those in (1) (multi-zone leptonic models). In hadronic processes, the high energy emission is generally due to proton synchrotron radiation and neutral pion decay. In addition, lepto-hadronic models that include emission processes from both theoretical approaches have been developed \citep[see e.g.][]{Bot19, Hov19, Cer20}. 

Depending on the strengths of optical/UV emission lines and relativistic beaming, blazars are predominantly classified as flat spectrum radio quasars (FSRQs), which are characterized by broad optical emission lines, or as BL Lacertae objects (BL Lacs), which generally lack strong emission or absorption features. On the basis of their synchrotron emission peak frequencies $\nu_{\rm peak}^{\rm sy}$, BL Lacs are further classified into LBLs\footnote{low-frequency peaked BL Lacs, $\nu_{\rm peak}^{\rm sy}$ < 10$^{14}$ Hz}, IBLs\footnote{intermediate-frequency peaked BL Lacs, 10$^{14}$ Hz < $\nu_{\rm peak}^{\rm sy}$ < 10$^{15}$ Hz}, HBLs\footnote{high-frequency peaked BL Lacs, 10$^{15}$ Hz < $\nu_{\rm peak}^{\rm sy}$ < 10$^{17}$ Hz} and EHBLs\footnote{extreme high-frequency peaked BL Lacs, $\nu_{\rm peak}^{\rm sy}$ > 10$^{17}$ Hz} \citep{Pado95,Cost01}. \cite{Abd10} devised a slightly different classification scheme, used in the Third Fermi High Energy Catalogue \citep[3FHL,][]{Ajello17}, which divides blazars into low-synchrotron peaked (LSP), intermediate-synchrotron peaked (ISP) and high-synchrotron peaked (HSP) classes. \cite{Hervet16} proposed a classification scheme that is based on the kinematics of the radio jets of blazars observed in very long base line interferometry. Their kinematic classification resulted in three classes of blazars: class I, class II and class I/II; they found the characteristics of their class I blazars to have a good correspondence to HBLs, class II overlaps the characteristics of FSRQs and their class I/II overlaps the characteristics of LBLs and IBLs. 

\smallskip




\cite{Goldoni21}--the predecessor of this paper and hereafter `\href{http://doi.org/10.1051/0004-6361/202040090}{Paper~I}'--provides a comprehensive introduction that is directly relevant to this work. We summarise in the next few paragraphs the main points of that introduction, highlighting the importance of optical spectroscopic observations of blazars. About 83~per cent of the approximately 4500 HE sources with a lower energy counterpart in the \emph{Fermi}-LAT \citep{Atwood2009} 4FGL-DR3 catalogue \citep{Abdo22} are associated with blazars, 20~per cent of which are FSRQs, 39~per cent are BL Lacs and the rest are blazar candidates of uncertain type (BCUs). Detections at VHE by existing Imaging Atmospheric Cherenkov Telescopes (IACTs; H.E.S.S.\footnote{\url{https://www.mpi-hd.mpg.de/hfm/HESS}}~\citep{hess1}, MAGIC\footnote{\url{https://magic.mpp.mpg.de}} \citep{magic1}, VERITAS\footnote{\url{https://veritas.sao.arizona.edu}} \citep{veritas1}) are rather limited, accounting for 81 blazars, 69 of which are BL~Lacs\footnote{\url{http://tevcat.uchicago.edu}}. 

The advent of the Cherenkov Telescope Array (CTA\footnote{\url{https://www.cta-observatory.org}}) will provide a substantial increase in our capacity to study the VHE Universe. CTA will have all-sky coverage with two sites, in the Northern (La Palma, Canary Islands, Spain) and Southern (Atacama desert, Chile) Hemispheres. With an energy range of 20 GeV - 300 TeV, excellent angular resolution and an order of magnitude increase in  sensitivity compared to existing IACTs, CTA is set to detect and study numerous amounts of distant VHE blazars. This will enable a more comprehensive population study of these objects across cosmic time, an important area of investigation that presently suffers from low number statistics at $z~\ge$ 0.2 \citep{Pita14}. Moreover, the increased detections will provide a window to investigate the density of extragalactic background light \citep[EBL,][]{Hau01,Bit15} as well as alternative physical processes that lead to the production of VHE gamma-ray emission. The propagation of this radiation allows us to investigate the intergalactic magnetic field properties, independently measure the Hubble constant $H_{\rm 0}$, and investigate possibilities of the existence of axion-like particles and search for Lorentz invariance violation, among other topics in beyond-standard-model physics \citep{Scta19}. 

CTA is poised to advance the current investigative efforts in all these areas \citep[see][and references therein]{cta21}. The first evidence of neutrino emission from BL Lacs \citep{IceCube18b,IceCube18a} provides another motivation for measuring the redshifts of these objects. A full understanding of the role that hadrons (and leptons) play in the jets of BL Lacs depends on the knowledge of total luminosity \citep[see e.g.][]{Cer15}, a quantity that is precisely estimated with redshift information \citep[see e.g.][]{Pai18}. Efforts aimed at gaining a better understanding of such unknowns are hampered by the difficulty involved in measuring BL Lac redshifts reliably, as their continuum-dominated optical spectra are nearly featureless, with their emission line equivalent widths (EWs) limit generally less than 5~\AA~\citep[][]{UrryPado95}, although \citet{Stick91} reported values higher than this. Attempts to mitigate this difficulty include taking high S/N optical spectra but even then, this approach is not always successful in detecting features in the resultant spectra. This has rendered a large fraction of BL Lac objects lacking redshifts to this day. 

Spectroscopic observational campaigns of \emph{Fermi}-LAT detected BL Lacs have been conducted by various groups, including \citet{Shaw13}, who observed 372 BL Lacs and, after combining their observations with those in the literature, the authors obtained redshifts for 44~per cent of their combined sample with a median redshift of $z_{\rm med}$ = 0.33. Other groups include \citet{Ace13,Arsi15,Arsi17,Kau19}, with some pursuing high S/N observations \citep{Pai17a, Lan18, Pai20} and others (e.g. \citet{Mas13,Pag14}) conducting observations of BCUs in low- and medium-sensitivity modes from locations both in the north and south. Recently, \citet{Pena20} analyzed a total of 416 observations of BL Lacs and BCUs, with the bulk of those coming from their own campaigns and the rest from the literature. The authors determined redshifts for about 30~per cent of their object sample with $z_{\rm med}$ = 0.285. 

Without precise redshift information, it is not possible to obtain source luminosities and this challenges our understanding of the blazar sequence--the notion that the peak frequency of a blazar SED becomes smaller with increasing bolometric luminosity $L_{\rm bol}$ \citep{Ghis17}. \citet{Giom15} postulate that BL Lacs lacking redshifts are mostly HSP objects with high luminosities. Measurements of more BL Lac redshifts and thus their luminosities could therefore provide a way to evaluate these two perspectives.

Because blazars are among the main targets for CTA, their true spectroscopic redshifts are earnestly sought. Such efforts are acknowledged to play an important supporting role in blazar science featured in the Key Science Program (KSP) on AGN for CTA \citep{Scta19}. It is with this in mind that we initiated spectroscopic observing campaigns to measure the redshifts of blazar candidates that CTA will most likely detect. We report the results of such observations in this second instalment in our series of papers to be published reporting the findings of our ongoing blazar redshift measurement efforts.

The paper is organized as follows: we present the sample selection, observing strategy, and observations and data reduction in Sections 2, 3 and 4, respectively. In Sections 5 and 6 we present the analysis and results and then provide the discussion and conclusions in Section 7. We used a cosmology with $\Omega_{M}$ = 0.308, $\Omega_{\Lambda}$ = 0.690 \citep{desCosmoParams2019} and $H_0$ = 67.8 km s$^{-1}$ Mpc$^{-1}$ \citep{desH02019} for all our calculations. All wavelengths and magnitudes are in air and in the AB system, respectively.

\section{Sample Selection}
The sample selection described in \href{http://doi.org/10.1051/0004-6361/202040090}{Paper~I} also applies to this work. Our focus is on the 1040 BL Lacs and BCUs from the 3FHL ($E$~>~10~GeV) \emph{Fermi}-LAT catalogue \citep{Ajello17}, 64~per cent of which have no redshift values in that catalogue. The minimum observational time requirement to detect each of such sources at  5$\sigma$ with the CTA array was estimated by performing Monte Carlo (MC) simulations with the Gammapy\footnote{\url{https://gammapy.org}} software \citep{gammapy:2017,gammapy:2019}, using the publicly-accessible CTA performance files\footnote{\url{https://www.cta-observatory.org/wp-content/uploads/2019/04/CTA-Performance-prod3b-v2-FITS.tar.gz}}$^{, }$\footnote{\url{https://zenodo.org/record/5163273\#.Yg9-yPVBzPZ}}. 
An extrapolation to very high energies of the reported average spectrum for each source in the 3FHL catalog was made. In order to simulate the expected spectral curvature at such energies, an \emph{ad hoc} exponential cut-off at 3 TeV is incorporated into the spectral model to ensure a more conservative modelling at the highest energies.

We use the optical depth $\tau$($E$,$z$) by \citet{Dom11} to account for the energy- and redshift-dependent effects of the EBL, where $E$ and $z$ are the gamma-ray energy and 3FHL source redshift, respectively. In doing that, a fixed redshift value of $z_{\rm fix}$ = 0.3--akin to  $z_{\rm med}$ = 0.33 of \citet{Shaw13} and $z_{\rm med}$ = 0.285 of \citet{Pena20} for BL Lacs--was assigned to sources with no reported redshift in 3FHL. While we are aware that we are biasing the sample with this selection, we choose this method for its simplicity.

In summary, the MC simulations resulted in 221 sources that CTA can detect in under 50 hours of observations. After revising 32 3FHL redshifts (see \href{http://doi.org/10.1051/0004-6361/202040090}{Paper~I} and its Appendix A for details), the MC simulations were repeated on the 221 sources, resulting in a selection of 165 sources with no redshift values that CTA is expected to detect in under 30 hours if found in their 3FHL flux state or in less time, if in a flaring state. In the following section, we explain the criteria for selecting sources from the sample for observations. 

\section{Observing Strategy}
The observing strategy of this paper follows directly from that of \href{http://doi.org/10.1051/0004-6361/202040090}{Paper~I}. Our main objective is to determine spectroscopic redshifts or lower limits for as many BL Lacs in our sample as possible.

In the observed spectra, we search for such stellar absorption features as the Ca \textsc{ii} HK doublet, Mg$_b$ and Na \textsc{i} D, which we expect from the luminous elliptical galaxies that are usually found to be the hosts of BL Lacs \citep{Urr00}. We also search for emission lines such as [O \textsc{ii}], [O \textsc{iii}], H$\upalpha$ and [N \textsc{ii}], but these are seldom detectable in the spectra. To ensure that we can measure EWs $\lesssim$~5~\AA, we require (1) that  the spectral resolution $\lambda$/$\delta\lambda$ has to be of the order of a few hundreds ($\geq$ 1000 is best), and (2) the S/N per pixel on average has to be of the order of 100. These two requirements combined are powerful enough to detect weak host galaxy features and absorption systems even with EW measurements lower than 5~\AA~\citep{Pita14}. If both these requirements are not possible in a spectrum, we configure our observing instruments in such a way that at least one of them is achievable. 

A thorough literature search was conducted for previous spectroscopic observations of the sources and evidence of their extended profiles arising from the host galaxies. Archival and published data sources such as the Two Micron All Sky Survey Extended Source Catalogue \citep{Skr06} were used in the search, resulting in the sources being classified as high-priority or low-priority. The former class is characterised by the availability of reliable information such as low S/N spectra and a tentative redshift. We present the observational results of such targets in this paper. The latter class, on the other hand, comprises sources with no extended profiles or at least one deep spectrum that is featureless. Efforts have been made by the authors to observe such sources when in their optical low states to benefit from the boosted spectral S/N of the host galaxy features that results from the reduced non-thermal emission (see \href{http://doi.org/10.1051/0004-6361/202040090}{Paper~I} for futher details).

\section{Observations and data reduction}

Spectral observations of the 33 blazars presented in this work were conducted in a similar manner as described in \href{http://doi.org/10.1051/0004-6361/202040090}{Paper~I}. In addition to conducting observations with the Keck/ESI\footnote{\raggedright Echellette Spectrograph and Imager (ESI) on the Keck II telescope, \url{https://www.keckobservatory.org/about/telescopes-instrumentation}}~\citep{Shei12}, SALT/RSS\footnote{Robert Stobie Spectrograph (RSS) on the Southern African Large Telescope (SALT), \url{www.salt.ac.za/telescope}}~\citep{Burgh03} and NTT/EFOSC2\footnote{ESO (European Southern Observatory) Faint Object Spectrograph and Camera (EFOSC2) on the New Technology Telescope (NTT), \url{https://www.eso.org/public/teles-instr/lasilla/ntt}}~\citep{Buz84} instruments, observations of 13 of the total sources were performed with the KAST Double Spectrograph (herein Lick/KAST) on the Shane 3-meter telescope\footnote{\url{https://mthamilton.ucolick.org/public/tele_inst/3m/\#primary}} at Lick Observatory. Keck II and SALT have primary mirror diameters of 10 and 11~m, respectively, whereas the primary mirror diameter for the NTT is 3.5~m. The 33 sources were observed for 66.3 hours in total, between 2019 April and 2021 March. Table \ref{tabobs1} lists the 25 blazars we report in detail in this work, comprising spectra of high S/N or in which  spectral features were detected. Table \ref{tabobslick} contains the observational results of twelve blazars observed with Lick/KAST, four of which are also listed in Table \ref{tabobs1} and eight of which are not reported in detail in the paper due to their low S/N featureless spectra and in the case of PKS 1424+240, due to a mistake in the instrument configuration in which a gap was left between the blue and red part of the spectrum.

\begin{table*}
  \caption{\label{tabobs1} List of observed sources with redshift information or high S/N and parameters of the observations. Source names with a $^{\dagger}$ at the end are listed in the BZCAT catalogue \citep{bzcat15}.}
\resizebox{17.8cm}{!}{%
\begin{tabular}{lcccccclcclll}
\hline\hline
Source & 3FHL name &   4FGL Name & Source name  & Ext. & RA & Dec  &   Telescope/ &   Slit            & Start Time & Exp.  & Airm. & Seeing      \\
number &  &   &    &      &     &      &   Instrument &  (\arcsec)  &  UTC       &  (s)   &         & (\arcsec)      \\  
(0) & (1) & (2)& (3) & (4) & (5) & (6) & (7) & (8) & (9)  &(10) & (11) & (12) \\ 
\hline
1 & 3FHL J0015.7+5551 & 4FGL J0015.6+5551 & GB6 J0015+5551  & N &  00 15 40.2 & +55 51 45 & Lick/KAST & 2.0 & 2019-08-29 09:37:56 & 5400 & 1.08 & 2.2 \\

2 & 3FHL J0045.7+1217 & 4FGL J0045.7+1217 & GB6 J0045+1217  & N & 00 45 43.4 & +12 17 12 & Lick/KAST   & 2.0 & 2019-11-01 04:46:40 & 5400 & 1.14 & 1.9 \\

3 & 3FHL J0054.7-2456 & 4FGL J0054.7-2455 & FRBA J0054-2455$^{\dagger}$& N & 00 54 46.8 & -24 55 29 & SALT/RSS & 2.0 & 2020-12-15 20:28:45 & 2325 & 1.30 & 1.7 \\

4 & 3FHL J0316.2-6439 & 4FGL J0316.2-6437 & SUMSS J031614-643732$^{\dagger}$ & N & 03 16 14.7 & -64 39 23 & SALT/RSS & 2.0 & 2020-12-24 20:37:37 & 2400 & 1.21 & 1.0 \\

  &                 &                   &                                  &  &  & &  & & 2021-01-07 20:24:51 & 2400 & 1.25 & 1.3 \\

   &                &                    &                                 &   &  & &  & & 2021-01-30 20:24:51 & 2400 & 1.30 & 1.7 \\

5 & 3FHL J0338.5+1302 & 4FGL J0338.5+1302 & RX J0338.4+1302  & N & 03 38 29.3 & +13 02 15 & Lick/KAST & 2.0 & 2019-11-01 08:30:48 & 7200 & 1.11 & 2.2 \\

6 & 3FHL J0403.2-2428 & 4FGL J0403.5-2437 & SHBL J040324.5-242950$^{\dagger}$ & N & 04 03 41.7 & -24 44 08 & NTT/EFOSC2 & 1.5 & 2020-02-18 00:36:24 & 4750 & 1.20 & 1.2 \\
 
7 & 3FHL J0500.7-4911 & 4FGL J0500.6-4911 & SUMSS J050038-491214 & N & 05 00 38.7 & -49 12 16 & SALT/RSS & 2.0 & 2020-01-24 22:03:56 & 2250 & 1.26 & 1.9 \\
 
  &                &                   &                      &  &  &  & &  & 2020-01-29 21:41:19 & 2250 & 1.27 & 1.4 \\

8 & 3FHL J0600.3+1245 & 4FGL J0600.3+1244 & NVSS J060015+12434$^{\dagger}$ & Y & 06 00 15.0 & +12 43 43 & Lick/KAST & 2.0 & 2019-11-01 10:49:28 & 9000 & 1.13 & 2.8 \\

9 & 3FHL J0604.2-4816 & 4FGL J0604.1-4816 &  1ES0602-48.2$^{\dagger}$       & N & 06 04 08.6 & -48 17 25 & SALT/RSS & 2.0 & 2020-01-26 22:50:10 & 2250 & 1.24 & 1.3 \\

  &                 &                  &                               &  &  &  & & & 2021-03-02 20:48:04 & 2250 & 1.29 & 1.1 \\

   &                &                   &                              &  &  &  &  & & 2021-03-24 19:13:14 & 2250 & 1.28 & 1.0 \\

10 & 3FHL J0819.4-0756  & 4FGL J0819.4-0756 & 1RXS J081917.6-0756$^{\dagger}$ & N & 08 19 17.6 & -07 56 26 & NTT/EFOSC2 & 1.5 & 2020-02-18 02:48:07 & 6650 & 1.11 & 1.0 \\
 
11 & 3FHL J1037.6+5711 & 4FGL J1037.7+5711 & GB6 J1037+5711$^{\dagger}$ & N &   10 37 44.3 & +57 11 55 &  Lick/KAST &  2.0  &  2019-04-11 03:15:27 &  3400 &  1.12 & 2.0 \\
 
12 & 3FHL J1041.9-0558 & 4FGL J1041.9-0557 & PMN J1042-0558 & N & 10 42 04.3 & -05 58 17 & NTT/EFOSC2 & 1.5 & 2020-02-18 05:24:49 & 3800 & 1.10 & 1.0 \\

13 & 3FHL J1130.7-3137 & 4FGL J1130.5-3137 & NVSS J113046$-$31380 & Y & 11 30 46.1 & -31 38 08 & NTT/EFOSC2 & 1.5 & 2020-02-18 06:57:00 & 2850 & 1.15 & 1.0 \\

14 & 3FHL J1259.9-3749 & 4FGL J1259.8-3749 & NVSS J125949-37485 & N & 12 59 49.8 & -37 48 58 & NTT/EFOSC2 & 1.5 & 2020-02-18 07:58:16 & 3800 & 1.02 & 1.4 \\

  &                 &                   &                   &  &  &  &  & & 2020-05-16 23:11:03 & 1274  & 1.37 & 1.5 \\

   &                &                    &                   & &  &     &  &  & 2020-05-31 21:47:29 & 2250  & 1.29 & 2.0 \\

15 & 3FHL J1304.3-4353 & 4FGL J1304.3-4353 & 1RXS J130421.2-435$^{\dagger}$ & N & 13 04 21.0 & -43 53 10 & SALT/RSS & 2.0 & 2020-06-19 20:42:40 & 2100 & 1.30 & 1.5 \\

16  & 3FHL J1427.0+2348 & 4FGL J1427.0+2348 & PKS 1424+240$^{\dagger}$   & N  & 14 27 00.4 & +23 48 00  & Lick/KAST & 2.0 & 2019-04-11 07:24:17  & 900  & 1.13  & 1.9 \\

17 & 3FHL J1532.7-1319 & 4FGL J1532.7-1319 & TXS 1530-131 & N & 15 32 45.4 & -13 19 10 & SALT/RSS & 2.0 & 2020-07-15 20:33:47 & 2250 & 1.28 & 2.0 \\

  &                &                   &             &  &  &  &  &  & 2020-08-18 18:21:17 & 2250 & 1.27 & 1.8 \\

   &               &                    &             &  &  &  &  & & 2020-08-19 18:20:52 & 2250 & 1.28 & 1.6 \\

18 & 3FHL J1549.9-0659  & 4FGL J1549.8-0659 & NVSS J154952-065907 & N & 15 49 52.0 & -06 59 08 & Keck/ESI & 1.0 & 2020-07-23 06:17:19 & 7200 & 1.17 & 0.8 \\

19 & 3FHL J1719.3+1206 & 4FGL J1719.3+1205 & 1RXS J171921.2+120  & N & 17 19 21.5 & +12 07 22 & Keck/ESI & 1.0 &  2020-07-23 08:30:12 & 7200 & 1.11 & 0.8 \\

20 & 3FHL J1844.4+1547 & 4FGL J1844.4+1547 & NVSS J184425+15464  & N & 18 44 25.4 & +15 46 46 & Lick/KAST   & 2.0 & 2020-07-21 06:54:48 & 7200 & 1.09 & 1.7 \\

21 & 3FHL J1933.3+0726 & 4FGL J1933.3+0726 & 1RXSJ193320.3+072$^{\dagger}$ & Y? & 19 33 20.3 & +07 26 22 & Keck/ESI   & 1.0 & 2020-07-23 10:39:18 & 5400 & 1.10 & 0.9 \\
 
22 & 3FHL J2031.0+1936 & 4FGL J2030.9+1935 & RX J2030.8+1935  & N & 20 30 57.1 & +19 36 13 & Keck/ESI & 1.0 & 2020-07-23 12:15:22  & 7200 & 1.18 & 1.0 \\

23 & 3FHL J2146.5-1343 & 4FGL J2146.5-1344 & NVSS J214637-13435$^{\dagger}$ & N & 21 46 37.0 & -13 43 60 & SALT/RSS & 2.0 & 2020-09-07 18:54:48 & 2013 & 1.17 & 2.2 \\

  &                &                   &                                &  &  &  &  &  & 2020-09-08 18:39:37 & 2250 & 1.19 & 1.8 \\

24 & 3FHL J2245.9+1545 & 4FGL J2245.9+1544 & NVSS J224604+15443 & N & 22 46 05.0 & +15 44 35 & Keck/ESI & 1.0 & 2020-07-23 14:20:29 & 1800 & 1.06 & 0.9 \\
 
25 & 3FHL J2321.8-6437  & 4FGL J2321.7-6438 & PMN J2321-6438 & N & 23 21 42.2 & -64 38 07 & SALT/RSS & 2.0 & 2020-09-07 21:03:02 & 2250 & 1.21 & 2.2 \\

  &                  &                   &               &  &  &  &  & & 2020-09-08 00:46:35 & 2250 & 1.21 & 1.1 \\

\hline
\end{tabular}
}
\begin{flushleft}
{\bf Notes.} The columns contain: (0) Source number, (1) 3FHL name, (2) 4FGL name, (3) Source name, (4) Extension flag: if a source is classified as extended, it is flagged with a yes (Y); if it is not classified as extended, it is flagged with a no (N), as discussed in Section 3, (5) Right ascension (J2000), (6) Declination (J2000), (7) Telescope and instrument, (8) Slit width, (9) Start time of the observations, (10) Exposure time, (11) Average airmass, and (12) Average seeing.
\end{flushleft}
\end{table*}

\href{http://doi.org/10.1051/0004-6361/202040090}{Paper~I} provides comprehensive technical and operational details of the Keck/ESI, SALT/RSS and NTT/EFOSC2 instruments.The Lick/KAST instrument has two spectrographs, where one is optimized for the blue end and the other for the red end, and has been operating at the Cassegrain focus of the Shane 3-m telescope since 1992\footnote{\url{https://mthamilton.ucolick.org/techdocs/instruments/kast}}. Each spectrograph has a throughput ranging between 5~per cent and $\sim$40~per cent, with the red end having a higher throughput\footnote{\url{https://mthamilton.ucolick.org/techdocs/instruments/kast/hw_detectors.html\#response}}. KAST observations were taken with the d55 dichroic, placing the division between the blue and red side at 5500 \AA. The 600/4310 grism was selected on the blue side and the 600/7500 grating was used on the red side.

In this work, we followed the same observational configurations, data reduction, order matching (applicable to ESI only), flux calibration, telluric correction and spectral dereddening procedures as those described in \href{http://doi.org/10.1051/0004-6361/202040090}{Paper~I} for observations made with the ESI, RSS and EFOSC2. Additionally for RSS,  while the PG0900 grating was used in longslit mode for the bulk of the observations, the higher resolution PG1300 grating was also used in longslit mode in observing the source PMN J2321-6438, after the initial observation with the PG0900 grating resulted in poorly resolved features. For EFOSC2, Grism 17 was used for the source SHBL J040324.5-242950, as its wavelength range (6895 - 8765 \AA) is better suited to the search of previously reported spectral features. For the KAST observations, data reduction and wavelength calibration were performed using the \textsc{iraf} software \citep{Tod86}. The \textsc{molecfit} program \citep{Sme15, Kau15} was used to perform telluric corrections and spectral dereddening was accomplished using maps by \citet{Sch11} (herein SF11) and the extinction curve by \citet{Fitz99}.

For flux calibration, one spectrophotometric standard star was taken for both ESI and EFOSC2 observations, while for the KAST observations, two spectrophotometric standard stars were observed typically at the beginning and at the end of each night. For the RSS observations, flux standards are not always included in the nightly observations and for that reason, we used proven good quality archival flux standards to perform flux calibration. Table \ref{tab_spgraphtech} provides a summary of the spectroscopic modes for the four observing instruments and the configurations while collecting observations for this work.

\begin{table*}
\caption{\label{tab_spgraphtech} Spectroscopic mode, wavelength coverage, throughput and spectral resolution of the four spectrographs, as used in this work.}

\centering
\begin{tabular}{lccccc}
\hline\hline

Instrument Name & Spectroscopic mode &  Wavelength coverage (\AA) & Throughput $p$ & Spectral resolution $\lambda$ / $\Delta \lambda$ \\

\hline

Keck/ESI  & Echellette & 3900 - 10000 & $p \geq$ 28\% & $\sim$ 10000 \\
SALT/RSS & Longslit & 4500 - 7500  & $p$ > 20\% & $\sim$ 1000 \\
NTT/EFOSC2 & Low resolution & 3860 - 8070  & 20\% < $p$ < 30\% & $\sim$ 500 \\
Lick/KAST & Blue channel & $\sim$ 3500 - 5600 & 5\% < $p$ < 20\% & $\sim$ 1000\\
Lick/KAST & Red channel & $\sim$ 5400 - 8000 & $p \sim$ 30\% - 40\% & $\sim$ 1500 \\

\hline
\end{tabular}
\end{table*}

\section{Redshift measurement and estimation of the blazar total emission}

Non-thermal emission from the jet and host galaxy \citep[usually elliptical,][]{Urr00} stellar emission give rise to the observed blazar SED in the optical regime. The much stronger jet emission overwhelms the host galaxy emission, rendering the host spectral features undetectable in most cases, as simulations by \citet{Landt02} and \citet{Pir07} have shown. Such simulations have also shown that host galaxy features are difficult to detect starting from a rest-frame jet-to-galaxy flux ratio (defined in the same way as in \href{http://doi.org/10.1051/0004-6361/202040090}{Paper~I)} of $\sim$10 at 5500 \AA.

Paper I describes in detail the steps involved in searching for redshift determination features in the spectra. Essentially, we thoroughly search for absorption and emission features--such as those given in Table \ref{TabLines}--in each spectrum. To measure a redshift convincingly, we require a minimum of two different features that yield the same redshift value. We then determine the EW of each line by normalising the spectrum with cubic splines and integrating the flux of each pixel. We estimate the EW measurement uncertainties from the error spectrum by taking the square root of its quadratic sum and by considering the continuum placement errors \citep[see][]{Sem92}. Tables \ref{tabeqw-abs} and \ref{tabeqw-emi} show the EW measured values.

\begin{table}
\caption{\label{TabLines} Absorption and emission features used in this work.}
\begin{tabular}{lll}
\hline\hline
Feature name & Wavelength (\AA) & Type \\
(1) & (2) & (3) \\
\hline
Ly~$\upalpha$ & 1215  & Absorption/Emission\\
Fe \sc{ii} & 2600 & Intervening\\
Mg \sc{ii} & 2796 & Intervening/Emission\\
       & 2803 & Intervening/Emission\\
{[O \sc{ii}]} & 3727 & Emission\\
       & 3729 & Emission\\
Ca \sc{ii} K & 3933.7 & Absorption\\
Ca \sc{ii} H & 3968.5 & Absorption\\
Ca \sc{i} G & 4304.4 & Absorption\\
H~$\upbeta$ & 4861.3 & Absorption/Emission\\
{[O \sc{iii}]} & 4959 & Emission\\
      & 5007 & Emission\\
Mg$_b$ & 5174 & Absorption\\
Ca Fe & 5269 & Absorption\\
Na \sc{i} D & 5892.5 & Absorption\\
{[N \sc{ii}]} & 6548.1  & Emission\\
H~$\upalpha$ & 6562.8 & Absorption/Emission\\
{[N \sc{ii}]} & 6583.6 & Emission\\
\hline
\end{tabular}
\begin{flushleft}
{\bf Notes.} The columns are: (1) Name of spectral feature, (2) Rest-frame wavelength in \AA, (3) Spectral feature type; whether it is a host galaxy emission or absorption feature or it is an intervening system. For features that are multiplets, e.g. Fe \textsc{ii} and Mg \textsc{ii}, we only list their strongest lines.
\end{flushleft}
\end{table}

The uncertainties on measured redshift values are estimated by taking into account the wavelength calibration uncertainties and the uncertainties of the positions at which the features are detected. In all our spectra, we have a wavelength calibration dispersion value of < 0.5~\AA~from roughly 4000~\AA~to $\sim$ 8000~\AA. This equates to a value lower than (6--12)~$\times~10^{-5}$ in relative precision. Gaussian functions were fitted at positions where the features were found for each source in Tables \ref{tabeqw-abs} and \ref{tabeqw-emi}. The variance of such fits was taken to be the uncertainty. The sums in quadrature of these two kinds of uncertainties are the total uncertainty estimates on the measured redshift values listed in Table \ref{tabres1}. 

After the redshift measurement steps, the source SED is modelled with a power law combined with elliptical galaxy templates \citep{Man01, Bru03}, where the former describes the jet emission and the latter the emission of the host. When needed, Gaussian emission features are added in the modelling  \citep{Pita14} and only one template is used per spectrum for simplicity. Our fitting process, performed with the \textsc{mpfit} software \citep{Mar09} involves two free parameters: the power law slope and jet-to-galaxy ratio. 
Table \ref{tabres1} shows the results of the fits.

Additionally, we estimate the absolute magnitudes of host galaxies that were detected. Slit losses are estimated by assuming the host galaxy effective radius $r_{\rm e}$ to be 10 kpc for a de Vaucouleurs profile. From the template spectra, we compute the K-corrections and do not apply evolutionary corrections. In the case of non-detection of a host, the spectra are fitted with a power law and normalised at the band centre. As the errors of the fitted parameters are unphysically small in such cases, we fit separate parts of the spectra to estimate them. The estimated host absolute magnitudes are presented in Table \ref{tabres1}.

\begin{table*}
\caption{\label{tabeqw-abs} Equivalent widths in \AA~of the absorption features detected in the spectra at the measured redshift.}

\centering
\begin{tabular}{lccccc}
\hline\hline

Source Name &  Ca \sc{ii} HK & Ca \sc{i} G & Mg$_b$ & Ca Fe & Na \sc{i} D  \\
            &       &      &     &       &     \\
 (1)  &  (2)   &  (3) & (4) & (5) &  (6)  \\        

\hline

GB6 J0015+5551  & 6.2$\pm$2.2   & -- &   2.5$\pm$0.9                 & 0.7$\pm$0.2                       &  4.8$\pm$0.3$^*$ \\
GB6 J0045+1217        & 2.9$\pm$0.7   & 2.6$\pm$0.7 & 3.8$\pm$0.8 & -- &  --   \\
SUMSS J031614-643732        & 2.1$\pm$0.4   & -- &  --  & --                        &  -- \\
SUMSS J0500-4912    & 6.6$\pm$0.8   & 1.3$\pm$0.2 &  --                   & --                      & -- \\
NVSS J060015+12434  & 7.1$\pm$3.8  & 4.4$\pm$0.8 
& 4.8$\pm$0.8  & -- & 1.3$\pm$0.2 \\

1ES 0602-48.2  & 2.4$\pm$0.3 & 1.0$\pm$0.2  &  -- &   -- &     -- \\

RX J0819.2-075    &    2.8$\pm$0.8  &  --  &  --  &  --  &   -- \\

PMN J1042-0558 &   3.4$\pm$0.6  &  --  &  --  &  --  &   -- \\

NVSS J113046-31380  &  10.4$\pm$0.8 & 3.9$\pm$0.7 & 8.5$\pm$0.8  &  2.2$\pm$0.4 & 8.6$\pm$0.6 \\

NVSS J125949-37485$^{**}$  &   4.2$\pm$0.5 & 0.6$\pm$0.2 &  3.6$\pm$0.4  &  -- &  -- \\

NVSS J125949-37485$^{***}$  &  2.2$\pm$0.8 &   1.5$\pm$0.3 & --      & --  &  -- \\

NVSS J154952-065907 & 2.5$\pm$0.3 & 0.7$\pm$0.1 & 0.7$\pm$0.1    & 0.3$\pm$0.1 & -- \\ 

RX J2030.8+1935  &  2.0$\pm$0.3 & 1.0$\pm$0.2$^*$ & 1.1$\pm$0.2 & -- & -- \\  

NVSS J224604+15443  &  10.7$\pm$1.3 & -- & -- & -- &    -- \\ 
\hline
\end{tabular}
\begin{flushleft}
{\bf Notes.} The columns contain: (1) Source name; Equivalent width with errors of the (2) Ca \textsc{ii} HK feature, (3) Ca \textsc{i} G feature, (4) Mg$_b$ feature, (5) Ca Fe feature, and (6) Na \textsc{i} D feature. If the feature is not detected, the legend is "--". The Na \textsc{i} D feature of GB6 J0015+5551 and the Ca \textsc{i} G feature of RXJ2030.8+1935, both flagged with asterisks, are likely contaminated by water absorption and Galactic Na \textsc{i} D, respectively. Two equivalent width results can be seen in the table for NVSS J125949-37485 as the source was observed by both SALT/RSS ($^{**}$) and NTT/EFOSC2 ($^{***}$).
\end{flushleft}
\end{table*}

\begin{table*}
\caption{\label{tabeqw-emi} Equivalent widths in \AA~of the main emission features detected in the spectra at the measured redshift.}

\centering
\begin{tabular}{lccccc}
\hline\hline

Source Name &  [O \sc{ii}] & [O \textsc{iii}]a & [O \textsc{iii}]b & H $\upalpha$ & [N \textsc{ii}]b \\
            &       &      &     &       &     \\
 (1)  &  (2)   &  (3) & (4) & (5) &  (6)  \\        

\hline

SHBL J040324.5-242950  & -- & 9.8$\pm$2.0 & 33.1$\pm$3.1 & -- & -- \\
SUMSS J0500-4912 & -- & -- & 1.4$\pm$0.2 & -- & --   \\
NVSS J060015+12434        & --   & -- &  --  & 0.7$\pm$0.2                        &  1.3$\pm$0.2 \\
RX J2030.8+1935 & 0.8$\pm$0.1   & -- &     0.7$\pm$0.1   & --                        & -- \\
NVSS J224604+15443  & 4.0$\pm$0.7  & --
& 3.7$\pm$0.5  & -- & -- \\

PMN J2321-6438  & 1.2$\pm$0.2 & -- &  -- &   -- & -- \\

\hline
\end{tabular}
\begin{flushleft}
{\bf Notes.} The columns are: (1) Source name; Equivalent width with errors of the (2) [O \textsc{ii}] feature, (3) [O \textsc{iii}]a feature, (4) [O \textsc{iii}]b feature, (5) H $\upalpha$ feature, and (6)~[N~\textsc{ii}]b feature. If the feature is not detected, the legend is ‘--’.
\end{flushleft}
\end{table*}

\begin{table*}
\caption{\label{tabres1} Analysis results on the spectra of all the observed sources.}
\centering
\begin{tabular}{lccccccc}
\hline\hline

 Source name  & S/N &  $R_{\rm c}$(BL Lac) & Redshift   & Flux Ratio  & $R_{\rm c}$(gal) & $M_{\rm R}$ & Slope   \\
              &     &   (obs)              &            &            &  (fit)            &   (gal)     &         \\  
   (1)  & (2) & (3)    &  (4)  &  (5)   &  (6)      &  (7)   &  (8) \\      
\hline
GB6 J0015+5551             & 19   & 16.3$\pm$0.1    &  0.2176$\pm$0.0004           &  1.1$\pm$0.2   & 16.9$\pm$0.3   &  -23.4   & -1.2$\pm$0.1  \\ 
GB6 J0045+1217             &  43  &  16.8$\pm$0.1  &  0.2544$\pm$0.0005            & 5.6$\pm$0.5    &  18.8$\pm$0.2  &  -22.0   & -1.2$\pm$0.1  \\
FRBA J0054-2455            & 101 &  16.8$\pm$0.2   &  --                                        &  --                   &  --                    &  --        & -1.1$\pm$0.1 \\
SUMSS J031614-6437     & 83   &  18.0$\pm$0.1   &  0.6161$\pm$0.0002           &  4.8$\pm$0.1   & 20.5$\pm$0.5  &   -23.4    & -1.3$\pm$0.2   \\ 
RX J0338.4+1302             &  32  & 16.6$\pm$0.2    & $\ge$ 0.3821$\pm$0.0002   &   --                  &  --                    &  --        & -1.7$\pm$0.1   \\      
SHBL J040324.5-242950 &  3    & 20.2$\pm$0.2    & 0.5993$\pm$0.0002             & --                   &  --                      &   --        &  -0.2$\pm$0.9   \\
SUMSS J0500-4912        &  50   & 18.3$\pm$0.1   &   0.2129$\pm$0.0001       & 2.9$\pm$0.2   &  19.8$\pm$0.1   &   -20.5     &  -0.7$\pm$0.1   \\
NVSS J060015+12434    & 25    & 15.7$\pm$0.2   &  0.0838$\pm$0.0003         &  0.7$\pm$0.1   & 15.9$\pm$0.2   &  -22.0     & -0.4$\pm$0.4    \\
1ES 0602-48.2                 & 72   &  17.9$\pm$0.1   & 0.4542$\pm$0.0002             &  6.6$\pm$0.4   & 20.0$\pm$0.3   &  -22.6     &  -1.2$\pm$0.1 \\
RX J0819.2-0756             & 36  & 18.4$\pm$0.1     &  0.320?$^*$                             &  1.7$\pm$0.2   & 19.5$\pm$0.3   &  -21.9   & -1.4$\pm$0.3  \\   
RX J0819.2-0756             & 36  & 18.4$\pm$0.1     &   --                                        &  --                  &  --                      &  --         & -0.6$\pm$0.1  \\
GB6 J1037+5711                 & 100    & 15.3$\pm$0.2 &   --                                      &   --                   &     --                  &    --       & -1.5$\pm$0.1  \\
PMN J1042-0558.            &   34   & 18.2$\pm$0.1  & 0.3925$\pm$0.0004             & 2.3$\pm$0.2    & 19.2$\pm$0.2    &  -22.9     & -1.8$\pm$0.1 \\
NVSS J113046-31380     &   40  & 17.4$\pm$0.1 &  0.1507$\pm$0.0003              &  0.3$\pm$0.1    &  17.4$\pm$0.1   & -22.0      & -1.3$\pm$0.3 \\
NVSS J125949-37485$\dagger$   & 48  & 17.3$\pm$0.1 & 0.2113$\pm$0.0006$^{**}$   & 2.2$\pm$0.7    & 18.3$\pm$0.2    & -22.0     & -0.9$\pm$0.1   \\
NVSS J125949-37485$\ddagger$ & 38  & 17.6$\pm$0.1 & 0.2107$\pm$0.0002$^{**}$ & 1.2$\pm$0.1    & 18.5$\pm$0.2    & -21.8 & -1.0$\pm$0.2   \\
1RXS J130421.2-435308  &  160  & 15.3$\pm$0.1   &  --                                      &   --                   &  --                     &  --         & -1.6$\pm$0.1    \\
PKS 1424+240 & 107 & 14.4$\pm$0.2 &  --                                      &   --                   &  --                     &  --         & -1.2$\pm$0.1 \\
TXS 1530-131                    & 5   & 22.3$\pm$0.2      &    --                                      &   --                  &    --                  &  --          & 2.6$\pm$0.1    \\
NVSS J154952-065907 & 101  & 16.9$\pm$0.1    &  0.4187$\pm$0.0005         &    4.0$\pm$0.4  & 18.5$\pm$0.2   &  -23.8     &  -1.3$\pm$0.1  \\
1RXS J171921.2+120711  & 85  & 17.6$\pm$0.2     &   --                                       &   --                  &    --                   &  --          & -0.8$\pm$0.2   \\ 
NVSS J184425+15464       &  50 & 15.8$\pm$0.2     &  $\ge$0.6293$\pm$0.0001 &  --                  &  --                     &  --         & -0.3$\pm$0.2 \\
1RXS J193320.3+072616  &   110 & 16.0$\pm$0.2   &      --                                   &  --                   &  --                     &  --           & -1.0$\pm$0.1    \\
RX J2030.8+1935               & 98   & 17.1$\pm$0.1    & 0.3665$\pm$0.0003      &  4.0$\pm$0.7   & 18.4$\pm$0.3.   &  -23.5      &  -1.2$\pm$0.2   \\
NVSS J214637-13435.       & 135  & 16.7$\pm$0.1   &   --                                    &   --                  &    --                    &  --           & -1.5$\pm$0.2   \\ 
NVSS J224604+15443       & 24  & 18.6$\pm$0.1     &  0.5966$\pm$0.0003     &  2.4$\pm$0.7 & 20.0$\pm$0.3.   &  -23.7      &  -0.6$\pm$0.1   \\
PMN J2321-6438                &  43    & 19.0$\pm$0.1 & 0.8126$\pm$0.0002           & --                    &  --                     &   --           &  -0.1$\pm$0.1   \\
PMN J2321-6438$\_{abs1}$ &  49   &  --                  & $\ge$ 0.7826$\pm$0.0003 & --                   &  --                     &   --           &   --   \\
PMN J2321-6438$\_{abs2}$ &  43    & --                  & $\ge$ 0.7901$\pm$0.0006  & --                  &  --                      &   --           &  --    \\
\hline
\end{tabular}
\begin{flushleft}
{\bf Notes:}  The columns contain: (1) Source name; (2) Median S/N ratio per spectral bin measured in continuum regions; (3) $R_{\rm c}$, Cousins magnitude of the BL Lac spectrum corrected for reddening, telluric absorption and slit losses with errors. Slit losses were estimated using an effective radius $r_e$ = 10 kpc for all sources for which a host galaxy was detected; (4) Redshift or lower limit with error; (5) Flux ratio jet/galaxy at 5500 \AA~in rest frame; (6) $R_{\rm c}$, Cousins magnitude of the galaxy with slit losses corrected assuming a 10 kpc radius as in column (3); (7) Absolute $R$ Magnitude of the galaxy - the errors are the same as those of column (6); (8) Power-law slope ($F_\lambda$~$\sim$~$\lambda^{\alpha_\lambda}$) with errors. If the entry is unknown, the legend is `--'. The possible redshift of RX J0819-0756 is based on a low confidence detection of the Ca \textsc{ii} HK feature. For this source, we also present the results of a simple power law fit. For NVSS J125949-37485, we report the results of the NTT/EFOSC2 and SALT/RSS observations separately. The results are compatible within errors taking into account the variability of the non-thermal component. The spectral bin width is 4 \AA~for the sources observed with EFOSC2, 1 \AA~for the sources observed with ESI and RSS and 3 \AA~for the sources observed with KAST.
    
$^{*}$ Uncertain redshift

$\dagger$ NTT/EFOSC2

$\ddagger$ SALT/RSS

$^{**}$ Combined redshift estimate is 0.2108 $\pm$ 0.002
\end{flushleft}
\end{table*}

\section{Sources and Results}
\label{sources}
In what follows, we discuss the observational results of each of the 25 sources that we report in detail.  Except otherwise stated, all our targets are classified as HBLs in the 3HSP catalog \citep{Chan19}. When the type of the source has not been published elsewhere, we examined its SED  as available online\footnote{\url{https://www.ssdc.asi.it/fermi4fgl}}. For four of these targets: GB6 J0015+5551, GB6 J0045+1217, NVSS J060015+124344 and NVSS J154952-065907, the redshift was published by other authors \citep{SDSSDR16, Pai20} after we performed our observations. Our results are compatible with theirs but we performed an additional analysis estimating the absolute magnitude of the host galaxy.

\subsection{GB6 J0015+5551}

 GB6 J0015+5551 is a BL Lac with an extended near-infrared (NIR) counterpart present in the 2MASX catalogue  \citep{Jar00}. A featureless low S/N spectrum was obtained at the Kitt Peak National Observatory using the R-C spectrograph \citep{Alv16} and confirmed its classification. It is located near the Galactic plane at longitude $b \sim -6.7^\circ$ and very absorbed with {\sl~E(B-V)}~=~0.37 \citep{Sch11}. Unless otherwise stated, the dust reddening ({\sl E(B-V)}) values quoted in the discussion of results for some of our sources below are those of \citet{Sch11}. In 2019 August, we observed it with Lick/KAST for 5400 s. The resulting spectrum is shown in Fig. \ref{fig_spec1}, top panel, on the left. At $z$ $\sim$ 0.217, Ca \textsc{ii} HK and Mg$_b$ are detected at slightly less than 3$\sigma$ while Na \textsc{i} D is detected at greater than 10$\sigma$ but is contaminated by water absorption. We note that the jet-to-galaxy flux ratio is quite low at 1.1 $\pm$ 0.2, which causes the absorption lines to be quite intense, allowing their detection even with a low-to-medium S/N spectrum. The resulting redshift is  $z$ = 0.2176 $\pm$ 0.0004, compatible with the one reported by \citet{Pai20}.

\subsection{GB6 J0045+1217}

 This BL Lac has a point-like optical counterpart in the SDSS database \citep{Bla17}. Inspection of its SED suggests that it is an IBL. Two featureless low signal-to-noise spectra have been reported  \citep{Shaw13,Krog15}. In 2019 November, GB6~J0045+1217 was observed with Lick/KAST, obtaining a medium S/N spectrum (Fig. \ref{fig_spec1}, top panel, on the right) in which we detected Ca \textsc{ii} HK, Ca \textsc{i} G and Mg$_b$ with low-to-medium signal-to-noise. We measure a redshift value $z$ = 0.2544 $\pm$ 0.0005. This is consistent with the values reported by \citet{Pai20} and \citet{SDSSDR16} and is a firm result.

\subsection{FRBA J0054-2455}

 Two medium S/N optical spectra of this BL Lac have been reported, one by \citet{Shaw13} using Keck/LRIS, the other by \citet{Mas13} using TNG/DOLORES, and both are featureless. We observed it with SALT/RSS for 2325 s in 2020 December obtaining S/N $\sim$ 100. After careful examination, a possible weak feature is detected around 4750 \AA\ in the integrated spectrum (Fig. \ref{fig_spec1}, middle panel, on the left). The putative feature is unresolved, its equivalent width is about 0.4 $\pm$ 0.2 \AA, at the sensitivity limit. No identification is possible at this stage. Given the low S/N of the feature, we consider this a tentative detection. Further observations are needed to investigate this result. The redshift of FRBA~J0054-2455 is still undetermined.

\subsection{SUMSS J031614-643732}

 This poorly investigated BL Lac has been observed by \citet{Lan15}. We performed three separate observations with SALT/RSS in 2020 December and 2021 January with a total exposure time of 7200 s. The total averaged spectrum is presented in Fig. \ref{fig_spec1}, middle panel, on the right. We detected the Ca \textsc{ii} HK doublet of the host galaxy at a significance of more than 5$\sigma$ at $z$ = 0.6161 $\pm$ 0.0002, making this one of the farthest blazars in our sample so far. The host galaxy is very luminous with $M_{\rm R}$ = -23.4.

\subsection{RX J0338.4+1302}

 The BL Lac RX J0338.4+1302 lies out of the Galactic plane at $b$~$\sim$~-33$^{\circ}$, but is heavily absorbed ({\sl E(B-V)} = 0.299, SF11) possibly due to the nearby Galactic dust cloud CODIR 174-34 \citep{Dutr02}. A Gran Telescopio Canarias\footnote{\url{http://www.gtc.iac.es/}} (GTC) spectrum \citep{Pai17a} resulted in the detection of an unresolved absorption feature with EW~=~3.0 \AA, which interpreted as Mg \textsc{ii}, indicates $z \ge$ 0.382. The feature is also visible in a previous spectrum by \citet{Marc16}. A Lick/KAST observation was performed for 7200 s. The resulting low S/N spectrum (Fig. \ref{fig_spec1}, bottom panel, on the left)  is a power law with a clear Mg \textsc{ii} absorbing doublet with EW~=~3.0~$\pm$~0.3~\AA~around 3870 \AA. We fit the feature with a Mg \textsc{ii} doublet using \textsc{vpfit} \citep{Cars14} and obtain a reduced $\chi^2 \sim 1.0$ for $z$ = 0.3821 $\pm$ 0.0002. We therefore determine that the target is at $z$ $\ge$ 0.3821. The ratio of the EW of the two components is 1.5, indicating a mildly saturated gas cloud.

\subsection{SHBL J040324.5-242950}

\begin{figure*}
  \centering
 \includegraphics[width=8.4truecm,height=6.955truecm]{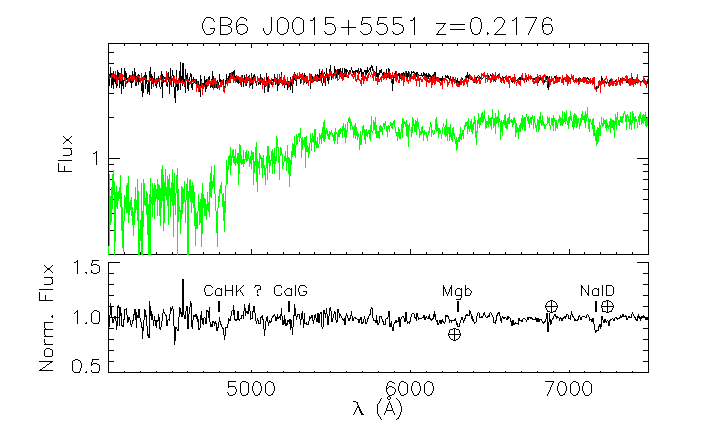}  \includegraphics[width=8.4truecm,height=6.955truecm]{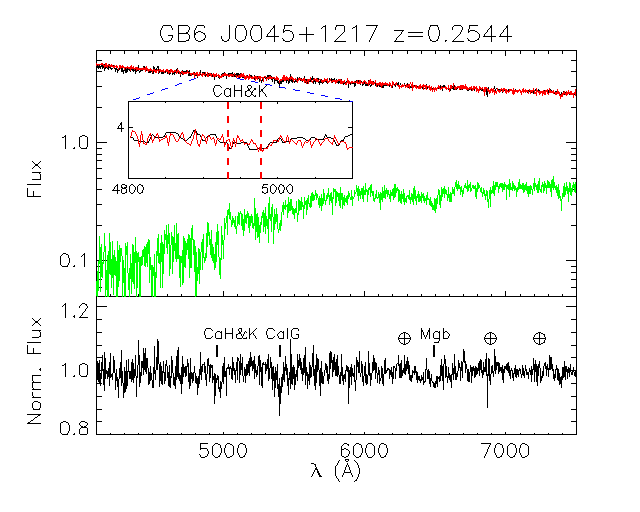} 
 \includegraphics[width=8.4truecm,height=7truecm]{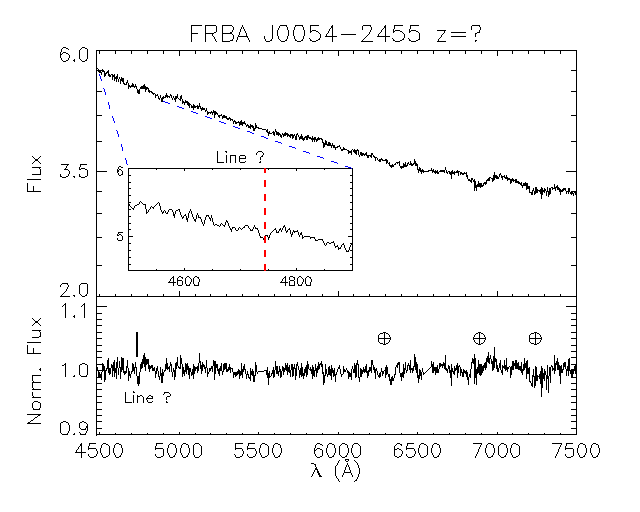}  \includegraphics[width=8.4truecm,height=7truecm]{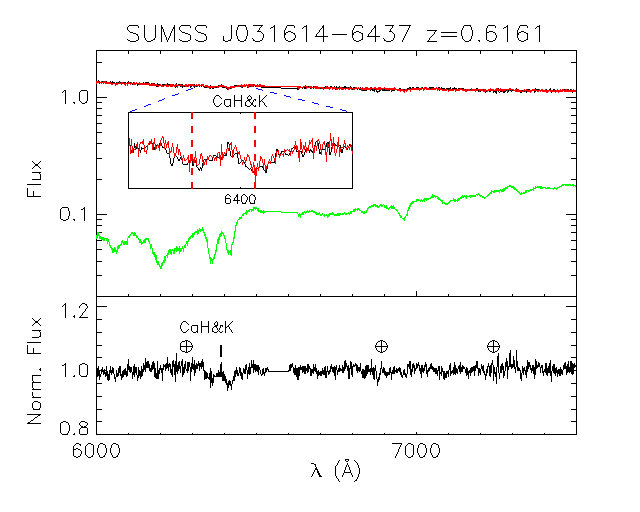}
  \includegraphics[width=8.4truecm,height=7truecm]{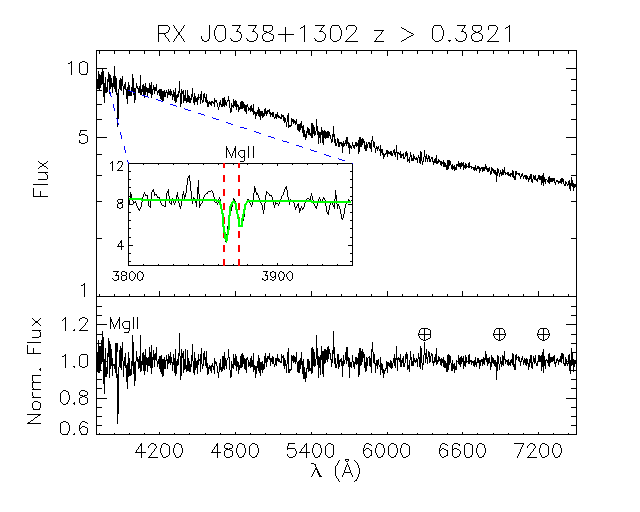}  \includegraphics[width=8.4truecm,height=7truecm]{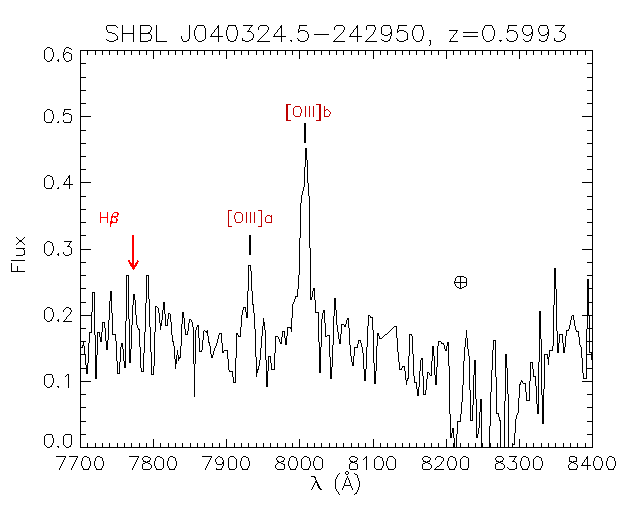}
  \caption{Spectra of the first six sources in Table \ref{tabobs1}. Each subfigure contains the spectrum, continuum, galaxy model for a given source and has an upper and lower panel, with the exception of the one for SHBL J040324.5-242950. {\sl Upper panel}: flux-calibrated and telluric-corrected spectrum (black) alongside the best-fitting jet+galaxy model when available (red). The flux is in units of 10$^{-16}$ erg  cm$^{-2}$ s$^{-1} $\AA$^{-1}$. The elliptical galaxy component is shown in green. {\sl Lower panel}: normalised spectrum with labels
for the detected absorption features. For SHBLJ 040324.5-242950, the observation was focused in the [O \textsc{iii}] doublet region and performed with a grism covering a limited wavelength range (see Sec. 6.6 for details). The flux is in the same units as in the other plots. Atmospheric telluric absorption features are indicated by the symbol $\oplus$ and Galactic absorption features are 
labelled `MW'. The flat or slanted regions such as those seen in the middle spectra are in general regions with bad telluric corrections and/or, for SALT/RSS spectra, they are also due to CCD gaps.}
 \label{fig_spec1}
    \end{figure*}

  SHBL J040324.5-242950 is classified as an HBL in the 3HSP catalog \citep{Chan19} with $\nu_{\rm peak}^{\rm sy}$ = 10$^{18}$ Hz. However, in the 4LAC catalog \citep{Fer4LAC20,Fer4LAC22}, $\nu_{\rm peak}^{\rm sy}$ = 5.5$\times$10$^{12}$ Hz is quoted, which would suggest it is an LBL. Two redshift values are reported in the literature for SHBL~J040324.5-242950: $z$ = 0.599 \citep{Heal08} and $z$~=~0.357 \citep{Giom05}. To investigate this result, we downloaded the public data of the observation by \citet{Heal08} from the ESO archive. The very low S/N spectrum displays a weak emission line around 8006 \AA~consistent with [O \textsc{iii}]b at $z$ = 0.599, supporting the result by \citet{Heal08}. In order to confirm or disprove this result, we observed SHBL J040324.5-242950 with NTT/EFOSC2 using Grism 17, which is sensitive in the range 6895~-~8765~\AA. Despite the non-ideal observing conditions, we were able to secure a low S/N spectrum. The spectrum presents a very faint continuum with S/N = 3 over which we clearly detect at redshift around 0.599 with 10$\sigma$ significance the [O \textsc{iii}]b line and with lower significance (4.6$\sigma$) the [O \textsc{iii}]a line (Fig. \ref{fig_spec1}, bottom panel, on the right). Only continuum flux was measured at the position of the H~$\upbeta$ line at this redshift, resulting in a 3$\sigma$ limit of an emission line of 9~\AA. The detected lines are both narrow and their flux ratio is [O~\textsc{iii}]b/[O\textsc{iii}]a~=~3.4~$\pm$~0.2, consistent with the expected value. The precise redshift value is $z$~=~0.5993~$\pm$~0.0002, confirming with higher precision the result by \citet{Heal08}. Finally, we note that the equivalent width of the [O \textsc{iii}]b line, EW = 33.1 +/-3.1 \AA~is much higher than the 5~\AA~limit used to separate BL Lacs and FSRQs and therefore suggest that the source might be an FSRQ. Interestingly, this possible classification would be consistent with the low $\nu_{\rm peak}^{\rm sy}$ quoted in the 4LAC catalog.

\subsection{SUMSS J050038-491214}

 This poorly studied BL Lac has no previous spectroscopic observation. Inspection of its SED suggests that it may be an HBL but not enough data is available. We observed it with SALT/RSS for 4500 s split in two separate observations both in 2020 January. On a power-law like spectrum, we detect the Ca \textsc{ii} HK doublet in absorption and [O \textsc{iii}]b in emission around $z$ = 0.212 (Fig. \ref{fig_spec2}, top panel, on the left). The precise redshift is $z$ = 0.2129 $\pm$ 0.0001. The host galaxy is particularly faint at $M_{\rm R}$ = -20.5.

\subsection{NVSS J060015+124344}

 NVSS J060015+124344 has an extended NIR counterpart in the 2MASX catalogue \citep{Jar00}. It is excessively absorbed with {\sl E(B-V)} = 0.41, being near the Galactic plane at $b \sim$ -5$^{\circ}$. We observed it in 2019 November for 7200 s with Lick/KAST. The spectrum in Fig. \ref{fig_spec2}, top panel, on the right, shows clearly the presence of host galaxy features and indicates that the source is a BL Lac. We detected Ca \textsc{ii} HK, Mg$_b$, Na \textsc{i} D in absorption and H $\upalpha$ and [N \textsc{ii}]b in emission, obtaining a precise redshift $z$ = 0.0838 $\pm$ 0.0003. This value agrees with that reported by \citet{Pai20}.
 
 \subsection{1ES 0602-48.2}

  A low S/N spectrum of this source was taken by \citet{Mas13} with NTT/EFOSC2. No clear feature could be detected. We performed three different observations with SALT/RSS for a total exposure time of 6750 s. In the averaged spectrum (S/N = 72) (Fig. \ref{fig_spec2}, middle panel, on the left), we detected Ca \textsc{ii} HK and Ca \textsc{i} G at $z$~=~0.4542~$\pm$~0.0002. The estimated host galaxy magnitude is $M_{\rm R}$ = -22.6.

\subsection{RX J0819.2-0756}

\begin{figure*}
  \centering
\includegraphics[width=8.4truecm,height=7truecm]{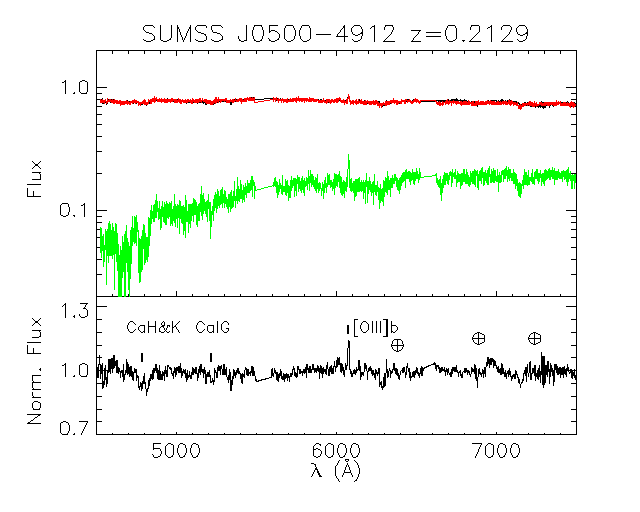}  \includegraphics[width=8.4truecm,height=7truecm]{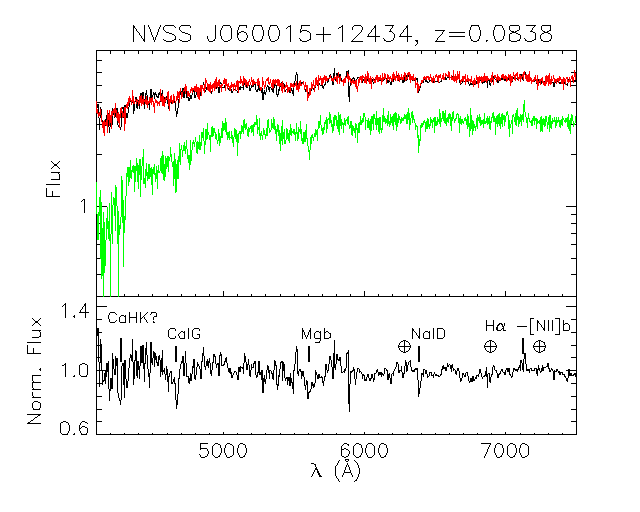}
 \includegraphics[width=8.4truecm,height=7truecm]{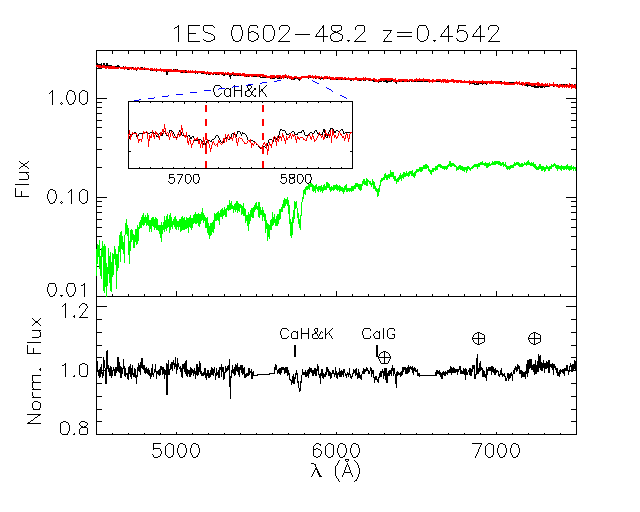}
 \includegraphics[width=8.4truecm,height=7truecm]{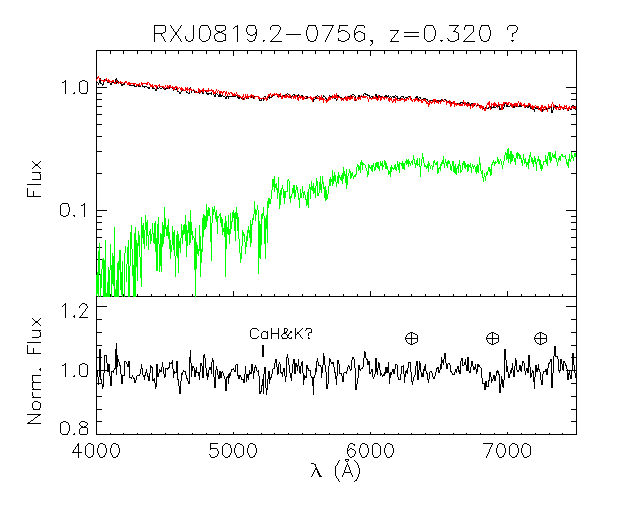} 
 \includegraphics[width=8.4truecm,height=7truecm]{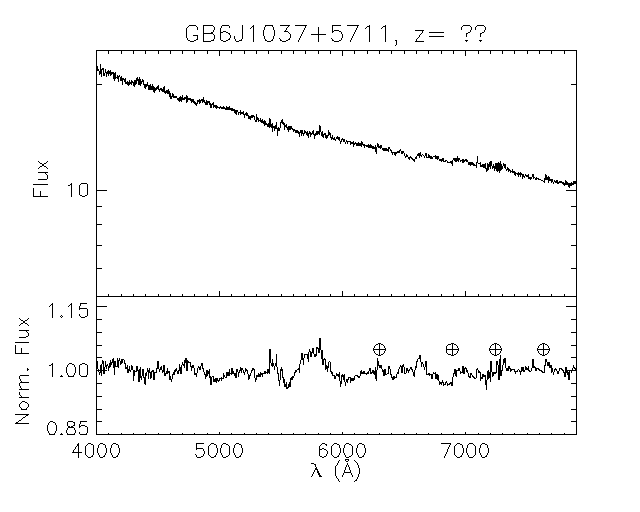}\includegraphics[width=8.4truecm,height=7truecm]{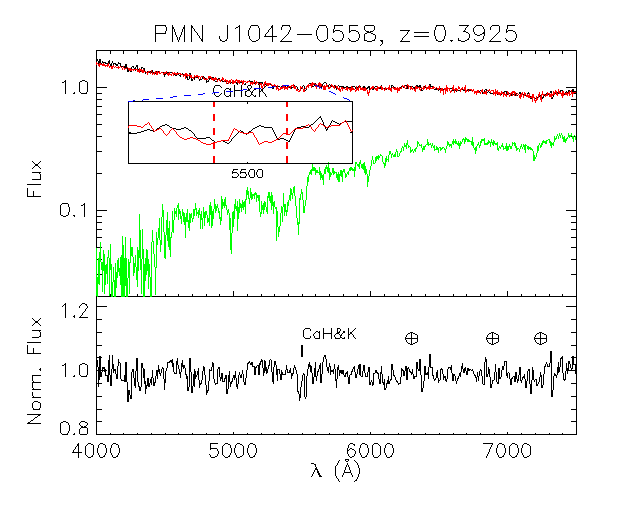}  

  \caption{Same as in Fig. \ref{fig_spec1} for sources 7 to 12 of Table \ref{tabobs1}.}
 \label{fig_spec2}
    \end{figure*}

  In the 6dF survey, it is claimed that this source is at $z$ = 0.85115 \citep{Jon09} while a photometric redshift $z$ = 0.45 has been estimated by \citet{Chan19}.  A featureless low S/N spectrum \citep{Alv16b} was obtained with a short (1200 s) observation using the IMACS medium resolution spectrograph at the 6.5 m Magellan telescope in Cerro Manqui, Chile. We observed the source with NTT/EFOSC2 for 6650 s, obtaining a moderate S/N spectrum. In this spectrum, presented in Fig. \ref{fig_spec2}, middle panel, on the right, we tentatively detected the Ca \textsc{ii} HK feature at 3.5$\sigma$ at $z \sim$ 0.320. Due to the weakness of the putative feature and the lack of other features, we consider this redshift tentative.
  
\subsection{GB6 J1037+5711}

This BL Lac has been observed in spectroscopy several times. Moderate S/N spectra  have been reported \citep{Shaw13,SDSSDR16} while a high S/N spectrum can be found in \citet{Pai20}. All these spectra are featureless and no redshift has been measured yet. A 3400 s observation was performed with Lick/KAST resulting in a S/N = 100 spectrum. No features could be detected (Fig. \ref{fig_spec2}, bottom panel, on the left). The redhisft of GB6 J1037+5711 remains undetermined.

\subsection{PMN J1042-0558}

PMN J1042-0558 has been classified as an EHBL \citep{Chan19}. A featureless low S/N spectrum has been obtained by \citet{Alv16b} with the Boller \& Chivens low resolution spectrograph at the Observatorio Astronomico Nacional 2.1 m telescope in San Pedro Martir (Mexico). We observed it with NTT/EFOSC2 for 3800~s, obtaining a spectrum with S/N = 34. In the spectrum (Fig.~\ref{fig_spec2}, bottom panel, on the right), we detected the Ca \textsc{ii} HK feature at $z$ $\sim$ 0.39 at 5.6$\sigma$. Other possible features could not be detected, as they fall on atmospheric absorption or they are out of the spectral range. Given the reliability of the detection of the Ca \textsc{ii} HK doublet however, we consider this a solid result. The resulting redshift of PMN J1042-0558 is $z$ = 0.3925 $\pm$ 0.0004.

\subsection{NVSS J113046-31380}

 NVSS J113046-31380 is an extended optical source found in the 2MASX catalogue \citep{Jar00}, with a possible redshift $z$~=~0.15 \citep{Jon09}. We obtained a medium S/N spectrum with NTT/EFOSC2, which is dominated by galaxy emission at $z~\sim$~0.15 with a weak power law component (Fig. \ref{fig_spec3}, top panel, on the left). Averaging on the detected features, we obtain $z$~=~0.1507~$\pm$~0.0003, confirming with higher precision the previous result.

\subsection{NVSS J125949-37485}

\begin{figure*}
  \centering
 \includegraphics[width=8.4truecm,height=7truecm]{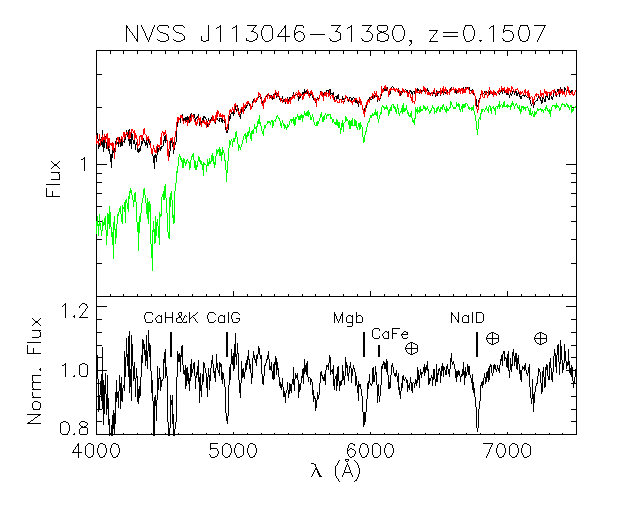}\includegraphics[width=8.4truecm,height=7truecm]{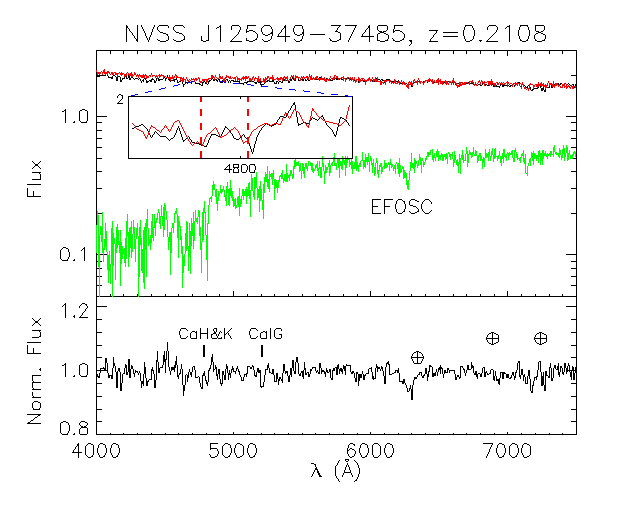}
 \includegraphics[width=8.4truecm,height=7truecm]{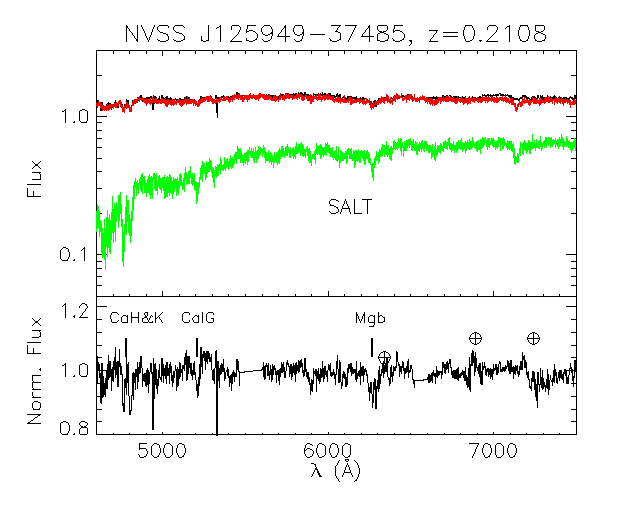}\includegraphics[width=8.4truecm,height=7truecm]{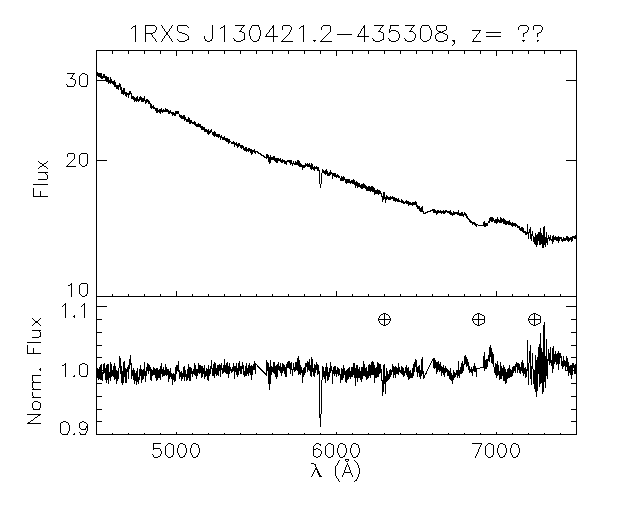} 
 \includegraphics[width=8.4truecm,height=7truecm]{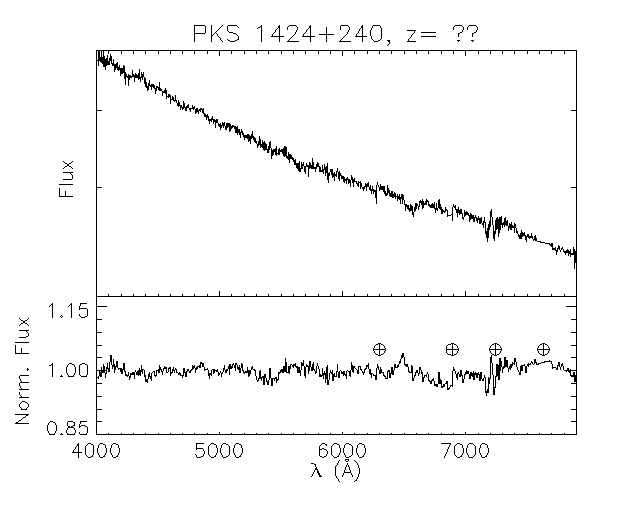}\includegraphics[width=8.4truecm,height=7truecm]{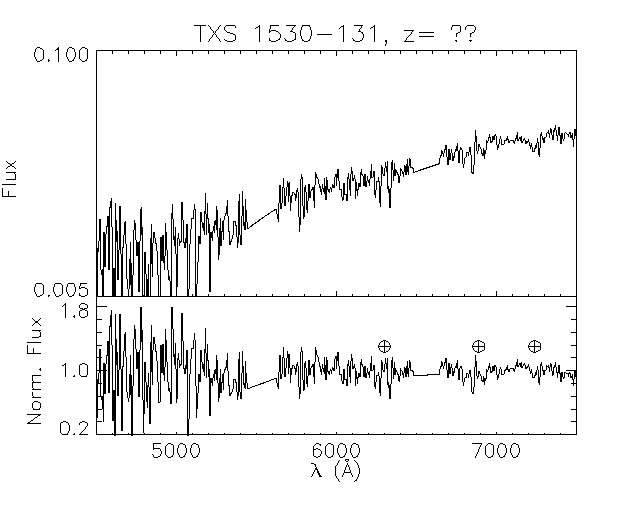}

  \caption{Same as in Fig. \ref{fig_spec1} for sources 13 to 17 of Table \ref{tabobs1}. Note that two spectra are present for NVSS J125949-37485, the redshift quoted in the title is the weighted mean of the redshifts quoted in Table \ref{tabres1}.}
 \label{fig_spec3}
    \end{figure*}

\cite{Ricci15} observed this BL Lac with the Goodman spectrograph on the SOAR telescope obtaining a low S/N spectrum without any detectable feature. We first observed the source with NTT/EFOSC2, obtaining a moderate S/N spectrum (Fig. \ref{fig_spec3}, top panel, on the right). In this spectrum, we tentatively detect at $z \sim$ 0.211 the Ca \textsc{ii} HK doublet at 3$\sigma$ and Ca \textsc{i} G at 5$\sigma$. The Mg$_b$ feature is possibly present but contaminated by atmospheric absorption. The observations were stopped due to bad weather and we could not confirm this result with further observations at NTT. From these observations, we estimated $z$ = 0.2113 $\pm$ 0.0006. In order to confirm or disprove this result, we observed the source with SALT/RSS and again secured a medium S/N spectrum (Fig. \ref{fig_spec3}, middle panel, on the left). In this case, however, we were able to detect Ca \textsc{ii} HK at 8$\sigma$ and Mg$_b$ at 9$\sigma$. We thus were able to determine the redshift of NVSS J125949-37485 as $z$ = 0.2107 $\pm$ 0.0002, fully compatible with but more precise than the NTT/EFOSC2 result. Despite the fact that the EFOSC2 and RSS spectra have similar continuum S/N, we think that  we were able to better detect the absorption features in the RSS spectrum due to a slight weakening of the non-thermal component. This is supported by the estimated magnitudes, which show that the source was brighter during the EFOSC2 observation (see Table \ref{tabres1}). The dimmer state in the RSS spectrum is also consistent with the larger equivalent width of Ca \textsc{ii} HK. Our best estimate of the redshift is 0.2108 $\pm$ 0.0002, based on the weighted average of the EFOSC2 and RSS results.

\subsection{1RXS J130421.2-43508}

This BL Lac was observed with NTT/EFOSC2 by \citet{Mas13}, obtaining a medium S/N featureless spectrum. We then observed it with SALT/RSS for 2100 s. The resulting spectrum with S/N $\sim$ 160 presented in Fig. \ref{fig_spec3}, middle panel, on the right, has no detectable spectral features. The redshift of 1RXS J130421.2-43508 thus remains unknown.

\subsection{PKS 1424+240}

This very bright TeV BL Lac has been sometimes classified as an ISP \citep{Nieppo06} and sometimes as an HSP \citep{Fer4LAC20}. Its redshift has been constrained to be $z \ge$ 0.603 with HST-COS detection of intervening Ly$\upalpha$ absorbers \citep{Fur13}. A galaxy group at $z$ = 0.601 has been possibly associated to it \citep{Rov16}. Moreover, the low significance detection of weak (EW~$\le$~0.1~\AA) emission lines at $z$ = 0.6047 \citep{Pai17a} in a very high S/N GTC spectrum has been reported. We observed PKS 1424+240 with Lick/KAST for 900 s to investigate this result. The resulting S/N = 107 spectrum  (see Fig. \ref{fig_spec3}, bottom panel, on the left) is featureless but does not have the sensitivity to detect the aforementioned lines. We cannot confirm the proposed redshift.

\subsection{TXS 1530-131}

TXS 1530-131 was first detected in gamma rays by \emph{Fermi}-LAT during a strong flare in 2011 \citep{Gasp11}. Its \emph{Fermi}-LAT light curve\footnote{\url{https://fermi.gsfc.nasa.gov/ssc/data/access/lat/msl_lc}} shows a strong variability. Its optical emission is quite weak \citep{Kau17} with $i^\prime$ = 18.70 $\pm$ 0.03 and no optical spectrum has been reported. We inspected its SED and found that the synchrotron peak is at around 10$^{13}$ Hz, suggesting that the source is an LBL. We observed TXS 1530-131 three different times with SALT/RSS during 2020 July and August for a total exposure time of 6750 s. The source was very weak with $R_{\rm c}$ = 22.3 $\pm$ 0.2, more than three magnitudes weaker than during the \citet{Kau17} observations. Given the weakness of the source, we could only obtain a very low S/N = 5 (Fig. \ref{fig_spec3}, bottom panel, on the right) spectrum. The redshift of TXS 1530-131 is still undetermined.

\subsection{NVSS J154952-065907}

  NVSS J154952-065907 was observed by \citet{Marc16} using the DOLORES spectrograph at the Telescopio Nazionale Galileo in the Canary Islands. They obtained a medium S/N spectrum with no lines. We observed this BL Lac with Keck/ESI for 7200 s, securing a spectrum with S/N of 101. We detect with high confidence Ca \textsc{ii} HK, Ca \textsc{i} G, Mg$_b$ and Ca Fe at $z$ = 0.4187 $\pm$ 0.0005 (Fig.  \ref{fig_spec4}, top panel, on the left).  This result is consistent with the one of \citet{Pai20}. The host galaxy has a high luminosity with $M_{\rm R}$ = -23.8.

\subsection{1RXS J171921.2+120711}

 1RXS J171921.2+120711 has a rather weak optical counterpart with $i$ = 18.3 in the PanSTARRS survey \citep{Mag20}. We could not find any optical spectrum of this BL Lac in the literature, and we therefore performed an exploratory observation with Lick/KAST. Only a low S/N spectrum could be obtained (see Table \ref{tabobslick}), which showed that a more powerful instrument was needed. We then took a spectrum with Keck/ESI, observing for 7200 s. In the resulting spectrum with S/N $\sim$ 85, no feature could be detected (Fig. \ref{fig_spec4}, top panel, on the right). The redshift of 1RXS J171921.2+120711 remains unknown.

\subsection{NVSS J184425+15464}

NVSS J184425+15464 is located near the Galactic plane at $b \sim$ +8.6$^{\circ}$ and is very absorbed with {\sl E(B-V)} = 0.3914. Two low S/N spectra were taken \citep{Mas15a, Alv16b} with the Boller \& Chivens low resolution spectrograph at the Observatorio Astronomico Nacional 2.1 m telescope in San Pedro Martir (Mexico) but no features could be detected. A photometric redshift $z$ = 0.11 \citep{Chan19} has been reported. We then observed it with Lick/KAST for 7200 s. A low S/N spectrum was obtained (Fig. \ref{fig_spec4}, middle panel, on the left), consisting of a flat power law on which a Mg \textsc{ii} absorbing system with EW = 1.3 $\pm$ 0.2 \AA~around 4550 \AA~is visible. We fitted the feature with a Mg \textsc{ii} doublet using \textsc{vpfit} \citep{Cars14}, obtaining a reduced $\chi^2 \sim$ 1.0 for $z$~=~0.6293~$\pm$~0.0006.  We therefore determine that the target is at $z$~$\ge$~0.6293, in contradiction to the photometric redshift quoted above. The ratio of the two Mg \textsc{ii} components is about 1, which implies a heavily saturated absorption system.

\subsection{1RXS J193320.3+072616}
\label{subsecj1933}

\begin{figure*}
  \centering
 \includegraphics[width=8.4truecm,height=7truecm]{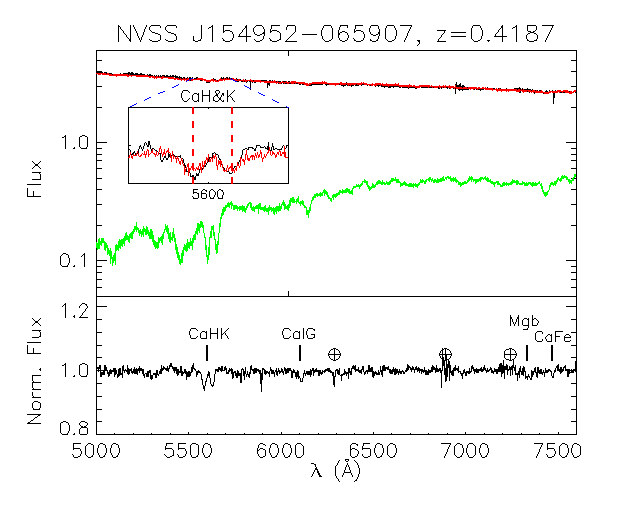}  \includegraphics[width=8.4truecm,height=7truecm]{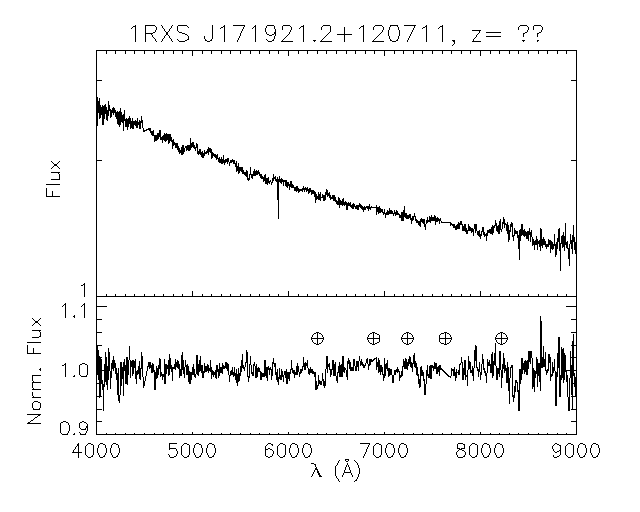}     
 \includegraphics[width=8.4truecm,height=7truecm]{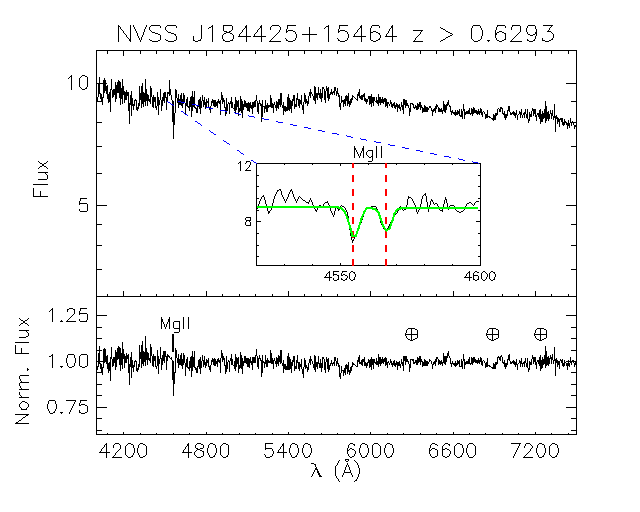}  \includegraphics[width=8.4truecm,height=7truecm]{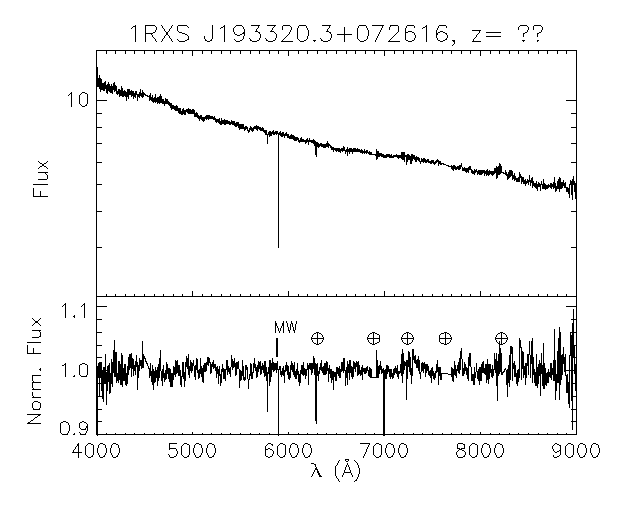}
  \includegraphics[width=8.4truecm,height=7truecm]{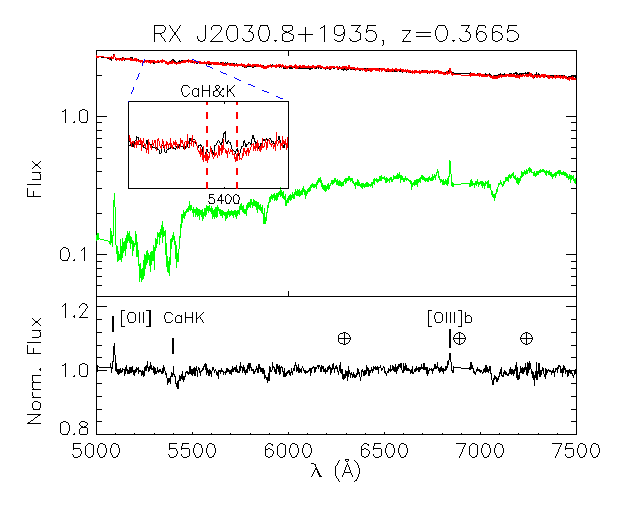}  \includegraphics[width=8.4truecm,height=7truecm]{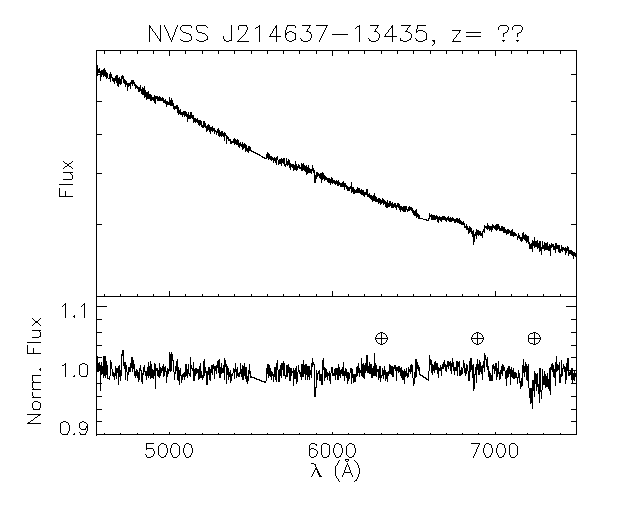}

 \caption{Same as in Fig. \ref{fig_spec1} for sources 18 to 23 of Table \ref{tabobs1}.  }
 \label{fig_spec4}
    \end{figure*}

A medium S/N featureless spectrum taken at Telescopio Nazionale Galileo with the DOLORES spectrograph of this heavily absorbed {\sl E(B-V)}~=~0.2539 source has been reported by \citet{Mas13}. Recent imaging observations have shown that it may be extended (Fallah Ramazani et. al in preparation). We performed two preliminary observations with Lick/KAST in 2019, obtaining medium S/N spectra (Table \ref{tabobslick}) but could not detect any feature. In 2020 July, a 5400 s observation was performed with Keck/ESI, resulting in a spectrum with S/N = 110 (Fig. \ref{fig_spec4}, middle panel, on the right). Unfortunately no extragalactic features could be detected in this spectrum and the redshift of 1RXS J193320.3+072616 remains undetermined. 
 
\begin{figure*}
  \centering
 \includegraphics[width=8.4truecm,height=7truecm]{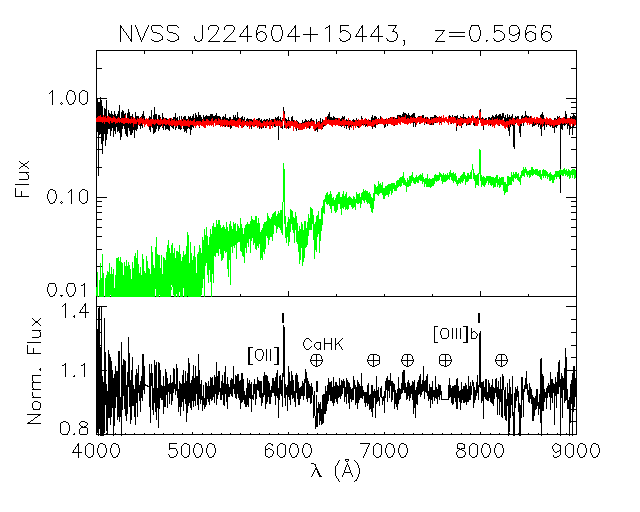}  \includegraphics[width=8.4truecm,height=7truecm]{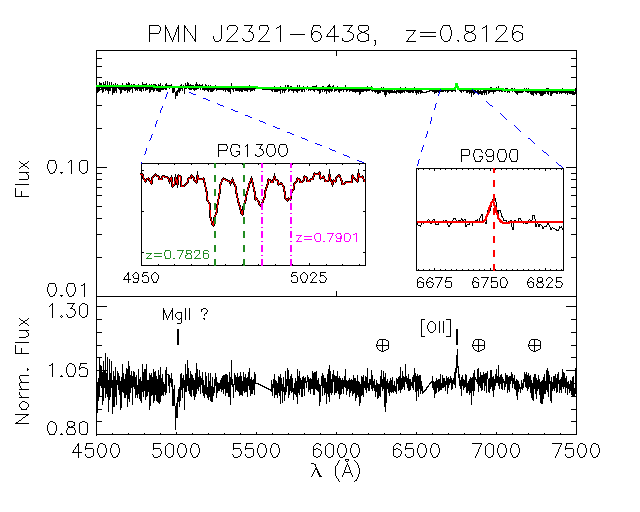}
\caption{Same as in Fig. \ref{fig_spec1} for sources 24 and 25 of Table \ref{tabobs1}.  }
 \label{fig_spec5}
    \end{figure*}

\subsection{RX J2030.8+1935}

\citet{Mas15a} reported a low S/N spectrum of this source, taken with the Boller \& Chivens low resolution spectrograph at the Observatorio Astronomico Nacional 2.1 m telescope in San Pedro Martir (Mexico). They tentatively detected a broad emission line around 4670 \AA, which, if interpreted as Mg \textsc{ii}, would put the source at $z \sim$ 0.668. In 2019, we performed an exploratory observation with Lick/KAST, obtaining a featureless spectrum with S/N = 12 (see Table \ref{tabobslick}). We thus could not confirm this feature. In order to obtain a high S/N spectrum, we then observed it with Keck/ESI for 2 hours obtaining a S/N = 98. In the resulting spectrum, we identify Ca \textsc{ii} HK, Ca \textsc{i} G, Mg$_b$, [O \textsc{ii}] and [O \textsc{iii}]b at $z$ = 0.3665 $\pm$ 0.0005 (Fig.~\ref{fig_spec4}, bottom panel, on the left). We thus contradict the interpretation of the  previous result. The host galaxy is quite luminous at $M_{\rm R}$ = -23.5.

\subsection{NVSS J214637-13435}

Two featureless medium S/N spectra \citep{Shaw13,Mas13} of this source have been reported. Furthermore, a very high photometric redshift, $z$ = 1.34 has been measured by \citet{Kau17}. This result is based on the detection of the Lyman $\upalpha$ break with multicolor {\sl Swift}/UVOT and GROND photometry. In order to investigate this interesting source, we performed two observations for a total observing time of 4263 s with SALT/RSS. No feature could be detected in the S/N = 135 spectrum and we could not determine the redshift of this source (Fig. \ref{fig_spec4}, bottom panel, on the right). Given that at $z$ = 1.34, Lyman $\upalpha$ is not detectable from the ground, we cannot confirm or disprove the photometric redshift.

\subsection{NVSS J224604+15443}

 This source is classified as a BCU in the 3FHL catalogue. Inspection of its SED suggests that it is an IBL. Two medium S/N spectra have been taken by \citet{Pai17a,Pai19} and both support its classification as a BL Lac.  In the first spectrum, a weak feature, contaminated by telluric absorption, is present around 6300 \AA. If interpreted as Ca \textsc{ii} HK, it implies $z \sim$ 0.6.  In the second spectrum, in addition to the contaminated feature, a weak [O \textsc{ii}] was also detected, leading to an estimated $z$ = 0.5965. We decided to perform a Keck/ESI observation to confirm or disprove this value. In a short 1800 s observation, we were able to detect [O \textsc{ii}], Ca \textsc{ii} HK and [O \textsc{iii}]b with high confidence at $z$ = 0.5966 $\pm$ 0.0003, confirming with higher precision the previous result (Fig. \ref{fig_spec5}, on the left).

\subsection{PMN J2321-6438}

PMN J2321-6438 is classified as a BCU in the 3FHL catalogue. It has been identified as a BL Lac with a low S/N featureless spectrum obtained by \citet{Des19} with the COSMOS spectrograph at the Blanco telescope in Chile. Inspection of its SED suggests that it is an IBL. We observed it with SALT/RSS the first time for 2250~s, obtaining a S/N = 43 spectrum. A single narrow ($\sim$ 450 km s$^{-1}$) emission line is clearly visible in the spectrum at around 6755 \AA, detected at 6$\sigma$ (Fig. \ref{fig_spec5}, on the right, right inset). The interpretation of this single line is ambiguous; usually in a BL Lac spectrum it could be [O \textsc{ii}] or [O \textsc{iii}]. If interpreted as [O \textsc{ii}], it would indicate $z~\sim$~0.812, and if interpreted as [O \textsc{iii}]b, it would indicate $z \sim$ 0.349. We also noticed a broad absorption feature around 5000 \AA~(EW $\sim$ 4 \AA). Such a feature would not be consistent with any host galaxy feature at the redshifts quoted above. However, it would be consistent with a strong Mg \textsc{ii} absorber around $z \sim$ 0.79 that would be unresolved using the SALT/RSS longslit PG0900 grating. We therefore performed a second observation using the higher resolution PG1300 grating. This grating has roughly double the resolution of the PG0900 grating ($\sim$~2000), in the spectral range 4650-6720 \AA. In this second spectrum, we clearly detect two Mg \textsc{ii} absorption systems: the bluer one at $z$~=~0.7826~$\pm$~0.0003 and the redder one at $z$ = 0.7901 $\pm$ 0.0006 (Fig.~\ref{fig_spec5}, on the right, left inset). The systems have respectively EW = 2.33 $\pm$ 0.15 \AA~and EW = 1.62 $\pm$ 0.14. The equivalent width ratio of the two components are respectively $\sim$ 1.4 and $\sim$ 1.1, indicating a moderately saturated system and a highly saturated one. These detections clearly establish that the source is at $z \ge$ 0.7901, and determines that the emission line previously detected is very likely [O \textsc{ii}]. The redshift of PMN J2321-6438 is thus $z$ = 0.8126 $\pm$ 0.0002. Additionally, the absence of bright emission lines confirms the \citet{Des19} identification of PMN J2321-6438 as a BL~Lac.

\section{Discussion and Conclusions}

In this work, 33 BL Lac objects, detected with {\sl Fermi}-LAT at energies $E\ge10$ GeV were observed with the following instruments: (1) the KAST double spectrograph on the Shane 3-m telescope at Lick observatory, (2) the EFOSC2 spectrograph on the NTT at the ESO observatory, (3) the RSS on SALT at the South African Astronomical Observatory and (4) the ESI spectrograph on Keck-II at the Keck Observatory. The spectra of 25 BL Lacs, containing spectral features or of higher S/N, have been presented in detail. The spectra for the remaining eight BL Lacs are of lower S/N and featureless and thus have not been presented in detail, but are briefly described in Appendix \ref{appA}. Our observing strategy was designed to have a spectral S/N ratio per pixel of 100 or more. Our aim was to measure the redshift or provide constraints, as well as the host galaxy properties for each BL Lac in our sample. We discuss in more detail in the following subsections the results of such measurements and provide a summary at the end.

\subsection{Spectral signal-to-noise ratio}

Out of the 25 BL Lacs with spectra containing spectral features or of higher quality, only seven of them reached our target S/N of $\ge$ 100: FRBA J0054-2455, GB6 J1037+5711, 1RXS J130421.2-43530, PKS 1424+240, NVSS J154952-065907, 1RXS J193320.3+07261, NVSS J214637-13435. Despite this, we were only able to measure the redshift for one of them. While in \href{http://doi.org/10.1051/0004-6361/202040090}{Paper~I} high S/N allowed us to get a higher efficiency, high S/N is not a guarantee for a redshift measurement, in particular during high activity periods of the non-thermal jet component. We note that, while the continuum spectrum of SHBL~J040324.5-242950 has a low S/N value of 3, the [O \textsc{iii}] emission line doublet is quite bright, which allowed its redshift determination. The S/N values for the remaining 17 BL Lacs are in the range 19 to 98, from which we measured 12 firm redshifts and one tentative value. We also measured two firm lower limits from the sample with S/N~<~100.

The six BL Lacs with spectral S/N~$\ge$~100 that did not result in a redshift measurement form part of the sample for which we will undertake further observations when the sources are in their optical low states. The authors have secured observing time on the Telescopi Joan Oró\footnote{\url{http://www.ieec.cat/content/18/generalities}} and Rapid Eye Mount \citep{Zer01} robotic telescopes to monitor the optical state of the sources and on SALT to perform spectroscopic ToO observations in low state. 

\subsection{Optical extensions of the sources}
As column 4 of Table \ref{tabobs1} shows, only two of our 25 sources reported in detail were found in the 2MASX catalogue \citep{Jar00} and had extended counterparts: NVSS J060015+124344 and NVSS~J113046-31380. The redshifts for both sources were measured (see Table \ref{tabres1}), supporting the claim from \href{http://doi.org/10.1051/0004-6361/202040090}{Paper~I} that sources with extended counterparts in the 2MASX catalogue are good candidates for redshift measurements. 1RXS J193320.3+072 has been found to have a possible extended counterpart by Fallah Ramazani et al. (in preparation) using imaging data; nevertheless its redshift could not be measured (see discussion in Section \ref{subsecj1933}).

\subsection{Host galaxy properties}
The fourteen firmly detected host galaxies in this work have an average magnitude of $M_{\rm R}$ = -22.6, which is similar to the value obtained in \href{http://doi.org/10.1051/0004-6361/202040090}{Paper~I} but with a greater dispersion of 1.0. While slightly more luminous, it is within the uncertainty limits of the value reported by \citet{Shaw13} but fainter than the values reported in \citet{Sbar05} and \citet{Pita14}. All 14 sources for which a host galaxy was detected in this work could be adequately fitted with a local elliptical template \citep{Man01}. Faint and narrow emission lines were detected for six of our objects. In two cases, SHBL J040324.5-242950 and PMN J2321-6438, the lines were decisive in the determination of the redshift as no absorption feature was detected. The equivalent width of the emission lines is smaller than the 5~\AA~limit traditionally used to separate BL Lacs and FSRQs, with the exception of SHBL J040324.5-242950, whose lines are much brighter.

\subsection{Comparison with Paper~I}
In Table \ref{resultsPaperI&II}, we compare the results of this work with those of Paper~I and the combined results. Overall, our redshift measurement efficiency for the 25 sources with good quality spectra that we report in detail in this work is about 56~per cent, which is roughly the same as in Paper~I. The  median redshift $z_{med}$ = 0.37 is higher compared to $z_{med}$ = 0.21 that we obtained in \href{http://doi.org/10.1051/0004-6361/202040090}{Paper~I}, implying a deeper redshift coverage in this work. The combined result of both Paper~I and Paper~II yields $z_{med}$ = 0.26 and a redshift detection efficiency of 57~per cent. The redshift detection efficiency for the seven targets that meet our S/N requirement of 100 or more turned out to be 14~per cent (1 out of 7) compared to 89~per cent (8 out of 9) that we obtained in Paper~I. 

\begin{table*}
\caption{\label{resultsPaperI&II} Number of observed sources, redshift and lower limit measurements for different groups of sources and for the whole sample for  \href{http://doi.org/10.1051/0004-6361/202040090}{Paper~I}, Paper~II (this paper) and the combined results (Paper~I + Paper~II)}.

\centering
\begin{tabular}{llllllllll}
\hline\hline

Paper & Number of & Redshifts & Redshift & Extended & S/N $\geq$ 100 & S/N < 100 & $z_{med}$ & Efficiency \\
& targets & ($z$) & lower limits & sources ($z$) & ($z$) & ($z$) & (with limits) & \\
(1)  & (2) & (3)    &  (4)  &  (5)   &  (6)      &  (7)   &  (8) & (9) \\  

\hline

I  & 19 & 11(+1) & 2(+1)& 8~~(7) & 9~~(8) & 10(3+1) & 0.21(0.23) & 11/19 ~~~~~~~| 58\%\\
II & 33 & 14(+1) & 2 & 3~~(2) & 7~~(1)& 26(13+1) & 0.37(0.38) & 14/25 (33) | 56\% (42\%) \\
I+II & 52 & 25(+2) & 4(+1) & 11(9) &  16(9) & 36(16+2) & 0.26(0.32) & 25/44 (52) | 57\% (48\%)\\
\hline
\end{tabular}

\begin{flushleft}
{\bf Notes:} The columns are: (1) Paper number; (2) Number of targets; (3) Number of redshifts measured; (4) Redshift lower limits; (5) Number of extended sources; (6) Number of sources with S/N $\geq$ 100; (7) Number of sources with S/N < 100; (8) Median redshift; (9) Redshift detection efficiency. The letter $z$ in brackets in columns 5 to 7 denotes the number of redshifts measured. In column 8, the values in brackets are the median redshifts with lower limits included in their calculations. The "+1" or "+2" notation in columns 3, 4 and 7 means an addition of one or two tentative redshifts or tentative lower limits to the total. In the last column, the survey efficiency of $\sim$ 42~per cent in brackets for Paper~II includes the addition of the eight sources with only low S/N spectra obtained with Lick/KAST (see Appendix \ref{appA}). The same description applies to the $\sim$ 48~per cent efficiency in brackets for the combined result. 
\end{flushleft}
\end{table*}

\subsection{Summary}
This is the second in a series of papers in which we report the results of our ongoing campaign to measure redshifts of blazars with a high probability of being detected by CTA. Our simulations have shown that the sources of our sample can be detected in less than 30 hours if observed in their average 3FHL state or in less time if observed in a flaring state. In particular, for the 25 sources with high S/N spectra we have discussed above in detail, the mean exposure time required for a detection in their average 3FHL state is 16.5 hours.

We list below the main results of this work:

\begin{enumerate}
    \item A total of 33 BL Lac objects were observed by four different telescopes. Twenty-five of them  either contain spectral features or have high S/N ratio featureless spectra and these we report in detail in the paper. The spectra for the remaining eight objects are featureless, and have low S/N ratio. We thus do not report them in detail but make reference to them in Appendix \ref{appA}. Twenty-two of the 25 BL Lacs had previous spectroscopic observations and 12 had unconfirmed redshifts in the literature. Our results confirm seven redshifts, contradict two and the other three remain unconfirmed. Overall, we measured 14 firm redshifts, one tentative redshift and two lower limits, all in the range 0.0838 $\leqslant z \leqslant$ 0.8126.
    
    \item Thirteen of the 25 sources were found to be at redshifts $z$ > 0.2, where the number of currently known VHE BL Lacs is fewer than 20. This resulted in a larger median redshift compared to \href{http://doi.org/10.1051/0004-6361/202040090}{Paper~I}.

    
    \item Compared to Paper~I, we achieved a low redshift measurement efficiency for high S/N spectra (1 out of 7) and a high efficiency for low S/N spectra (13 out of 25 or out of 33, if the eight sources referenced only in Appendix \ref{appA} are included).
    
    \item As with Paper~I, we achieved a roughly similar high redshift measurement efficiency for the two (or three - see Table \ref{tabobs1}) sources with extended optical/NIR counterparts.
    
    \item Our measured average host galaxy magnitude is $M_{\rm R}$ = -22.6, a value we also obtained in \href{http://doi.org/10.1051/0004-6361/202040090}{Paper~I}. We find a larger spread of 1.0 in comparison to 0.4 in Paper I. We again find the host galaxy properties to be consistent with those of ellipticals.
    
\end{enumerate}

Our observations in support of blazar science for the CTA Key Science Project on AGN are ongoing. Not only do these observations support the CTA science goals but also the science goals of the large astronomy community engaged in studies of blazars for which knowledge of redshifts is crucial. We will undertake re-observations under our existing and future programmes of sources that have S/N~<~100 in their spectra and in which spectral features could not be detected to attain our target S/N $\geq$ 100, thereby increasing the odds of determining their redshifts. Targets not falling in this category, i.e. having spectral S/N > 100 with no detection of features in the spectra, will only be re-observed when in a lower optical state. This technique has proven effective in \href{http://doi.org/10.1051/0004-6361/202040090}{Paper~I}, as it allowed the redshift of MAGIC J2001+435 to be measured. We have also recently observed the blazar PKS 1424+240 in an optical low state with OSIRIS on GTC in order to investigate in greater detail the proposed redshift (Becerra González et al. in preparation).

\section*{Acknowledgements}
Based on observations collected at the European Organisation for Astronomical Research in the Southern Hemisphere, Chile, under programs P104.B-0432(A). Based on observations made with the Southern African Large Telescope (SALT) under programs 2020-1-SCI-027 and 2020-2-SCI-040 (PI E. Kasai). This work was supported in part by the grants PHY-1707432 and PHY-2011420 from the U.S. National Science Foundation. Some of the data presented herein were obtained at the W. M. Keck Observatory, which is operated as a scientific partnership among the California Institute of Technology, the University of California and the National Aeronautics and Space Administration. The Observatory was made possible by the generous financial support of the W. M. Keck Foundation. The authors wish to recognise and acknowledge the very significant cultural role and reverence that the summit of Maunakea has always had within the indigenous Hawaiian community. We are most fortunate to have the opportunity to conduct observations from this mountain. We are grateful to the staff at Lick Observatory, the W. M. Keck Observatory, ESO and South African Astronomical Observatory (SAAO) for their outstanding and tireless support during our observing runs. W. Max-Moerbeck gratefully acknowledges support by the ANID BASAL projects ACE210002 and FB210003, and FONDECYT 11190853. This research made use of the CTA instrument response functions provided by the CTA Consortium and Observatory, see \url{https://www.cta-observatory.org/science/cta-performance} (version prod3b-v1) for more details. We gratefully acknowledge financial support from the agencies and organisations listed here: \url{http://www.cta-observatory.org/consortium\_acknowledgments}. Lastly, we thank the reviewer and editor for a constructive review of our manuscript. It greatly enhanced the reporting of our findings, especially in the discussion section. This research made use of the SIMBAD database, operated at CDS, Strasbourg, France.


\section*{Data Availability}

The raw FITS data files are available in the Lick, ESO, SAAO and Keck archives. The data underlying this article will be shared on reasonable request to the corresponding author.




\bibliographystyle{mnras}
\bibliography{paper}




\appendix

\section{Additional Observations with the Lick telescope}
\label{appA}

We report here on observations performed with Lick/KAST yielding S/N < 100 and no detection of spectral features. The observations are for a total of 12 blazars. Four of the blazars have higher quality spectra obtained with other instruments that are reported in Table \ref{tabobs1}.

\begin{table*}
    \caption{\label{tabobslick} Analysis results on 17 featureless spectra of twelve blazars observed with Lick/KAST. They include four spectra of sources reported in Table \ref{tabobs1} of this paper: PKS 1424+240, 1RXS J171921.2+120711, 1RXS J193320.3+072616 and RXJ2030.8+1935. The remaining 13 spectra are of eight sources not otherwise reported in detail in this paper. Note that the spectrum of PKS 1424+240 has formally S/N higher than our threshold. However, due to a mistake in the instrument configuration, a gap was left between the blue and red part of the spectrum. We thus list it in this table. Source names with a $^{\dagger}$ at the end are listed in the BZCAT catalogue \citep{bzcat15}.}
    \resizebox{17.8cm}{!}{%
        \begin{tabular}{lcccclccllccc}
          \hline\hline
            3FHL name &   4FGL name & Source name  & Ext. & RA & Dec   & Start time & Exp.  & Airm. & Seeing   & S/N & Slope & $R_{\rm c}$  \\
        
                      &                        &                         &        &       &           &     UTC     & (s)   &         &   (\arcsec) &.   &           &   \\
             (1) & (2) & (3) & (4) & (5) & (6) & (7) & (8) & (9)  & (10) & (11) & (12)  & (13)  \\ 
            \hline
            3FHL J0134.4+2638 & 4FGL J0134.5+2637  & 1RXS J013427.2+263846$^{\dagger}$ & N &  01 34 28.3  & +26 38 45 &  2019-11-02 06:23:41 & 7200 & 1.04 & 2.2 & 29 & -1.5$\pm$0.1 & 16.9$\pm$0.2 \\

            3FHL J1150.5+4154 & 4FGL J1150.6+4154 & RBS 1040$^{\dagger}$ & Y? & 11 50 34.8 & +41 54 40&  2019-04-11 04:38:16  & 7850 & 1.04 & 1.8 & 59 & -1.0$\pm$0.1& 16.6$\pm$0.2 \\

            &  &            &  &   & &  2020-05-26 04:35:42  & 7200 & 1.07 & 2.4 & 50 & -1.4$\pm$0.2 & 16.5$\pm$0.2 \\

            3FHL J1427.0+2348 & 4FGL J1427.0+2348 & PKS 1424+240$^{\dagger}$   & N &   14 27 00.4 & +23 48 00  & 2019-06-08 04:35:14 & 4500 & 1.04& 2.8 & 112 & -1.2$\pm$0.1 & 14.0$\pm$0.2 \\

            3FHL J1546.1+0818 & 4FGL J1546.0+0819 & 1RXS J154604.6+081912$^{\dagger}$ & N & 15 46 04.3  & +08 19 13 &  2019-04-11 08:22:01 & 9050 & 1.22 & 2.9 & 26 & -1.5$\pm$0.2 & 17.3$\pm$0.3\\

            3FHL J1719.3+1206 &4FGL J1719.3+1205 & 1RXS J171921.2+120711 & N & 17 19 21.5  & +12 07 22 &  2020-05-26 06:56:32  & 7020 & 1.19 & 2.4 & 12 & -0.6$\pm$0.3 & 19.0$\pm$0.1\\

            3FHL J1725.0+1152  &4FGL J1725.0+1152 & 1H 1720+117$^{\dagger}$   & N &  17 25 04.3  &  +11 52 15  & 2019-04-11 11:29:55  & 3600 & 1.12 & 2.9 & 43 & -0.7$\pm$0.5 & 15.6$\pm$0.1\\
        
            &&&&& & 2019-06-08 06:08:39  & 3600 & 1.32 & 3.3 & 33 & -1.0$\pm$0.1 & 15.6$\pm$0.1\\

            3FHL J1811.3+0341 &4FGL J1811.3+0340  & NVSS J181118+03411$^{\dagger}$ & N &  18 11 18.1 &  +03 41 14 &2020-05-26 09:42:05  & 7200 & 1.22 & 2.7 & 60 & -0.5$\pm$0.4 & 16.7$\pm$0.1\\

            &&&&& &2020-07-21 04:45:32  & 5400 & 1.22 & 2.0 & 50 & -1.2$\pm$0.2 & 15.5$\pm$0.2\\

            3FHL J1933.3+0726 & 4FGL J1933.3+0726&  1RXS J193320.3+072616$^{\dagger}$ & Y? & 19 33 20.3 & +07 26 22 & 2019-08-30 04:03:30  &  5400 &1.19  & 1.9 & 43 & -0.9$\pm$0.1 & 16.1$\pm$0.2\\

            3FHL J2031.0+1936 & 4FGL J2030.9+1935&  RX J2030.8+1935        &  N  & 20 30 57.1 &+19 36 13 &2019-11-02 02:33:25   &7200  &  1.13 & 2.1 & 22 & -0.8$\pm$0.1 & 17.1$\pm$0.2\\

            3FHL J2156.0+1818 & 4FGL J2156.0+1818 &  RX J2156.0+1818      &  N  &21 56 01.6  &+18 18 37& 2019-08-30 05:26:06  & 5400 &1.14  & 2.1 & 55&-1.3$\pm$0.1 & 16.6$\pm$0.3\\

            &&&&&& 2020-07-21 08:38:10  &  1800 &1.12 & 2.5 & 26 & -1.1$\pm$0.1 & 16.9$\pm$0.2\\

            3FHL J2247.9+4413 &4FGL J2247.8+4413  &  NVSS J224753+44131$^{\dagger}$ &N  &  22 47 53.2 & +44 13 15& 2020-07-21 09:15:02  &  7200 &1.05& 2.5&  44 &-1.2$\pm$0.1 & 16.3$\pm$0.2\\
            
            3FHL J2304.7+3705 &4FGL J2304.6+3704  & 1RXS J230437.1+370506$^{\dagger}$ & N & 23 04 36.7& +37 05 07 & 2019-08-30 07:32:45&  5400 &1.01 & 2.1&  33 & -1.5$\pm$0.1 & 17.1$\pm$0.2\\

            &&&&& & 2020-07-21 11:25:43 &  1800 &1.01 & 2.5&  16 & -1.4$\pm$0.1 & 17.2$\pm$0.2\\
            \hline
        \end{tabular}
    }
    \begin{flushleft}
            {\bf Notes.} The columns contain: (1) 3FHL name; (2) 4FGL name; (3) Source name; (4) Extension flag, as in Table \ref{tabobs1}; (5) Right ascension (J2000); (6) Declination (J2000); (7) Start time of the observations; (8) Exposure time; (9) Average airmass; (10) Average seeing; (11) Median S/N ratio per spectral bin measured in continuum regions; (12) Power-law slope with errors; and (13) $R_{\rm c}$, Cousins magnitude of the BL Lac spectrum corrected for reddening, telluric absorption and slit losses with errors.
    \end{flushleft}
\end{table*}



\bsp	
\label{lastpage}
\end{document}